\documentclass[aps,prl,superscriptaddress,twocolumn,floatfix]{revtex4-2}

\usepackage{bm}
\usepackage{physics}
\usepackage{graphicx}
\usepackage{float}
\usepackage{multirow}
\usepackage{amsmath,amssymb,amsfonts,amsthm,mathtools}
\usepackage[colorlinks=true,linkcolor=blue,citecolor=blue,urlcolor=blue]{hyperref}

\begin{document}

\title{Unified Theory for Momentum-Space Nonsymmorphic Symmetry of Spin Space Groups}

\author{Peiyuan Wang}
\affiliation{Department of Physics and HK Institute of Quantum Science \&
Technology, The University of Hong Kong, Pokfulam Road, Hong Kong, China}

\author{Zheng Zhang}
\affiliation{Department of Physics, School of Science, Lanzhou University of
Technology, Lanzhou 730050, China}

\author{Y. X. Zhao}
\email{yuxinphys@hku.hk}
\affiliation{Department of Physics and HK Institute of Quantum Science \&
Technology, The University of Hong Kong, Pokfulam Road, Hong Kong, China}

\begin{abstract}
Projective crystal symmetry provide a route to nonsymmorphic symmetry structures in momentum space, but a general theory for their realization in spin space groups has been lacking. 
We show that a nonsymmorphic momentum-space action is possible only when the unitary translation subgroup is combined with rotations about a common axis for a coplanar or noncoplanar spin space group of spin-\(1/2\) systems.  Its factor system originates entirely from the double cover \(\mathrm{Spin}(2)\to SO(2)\), yielding a general formula for the fractional reciprocal-lattice translations and hence for the point-group action on the Brillouin zone. We further prove that a momentum-space nonsymmorphic group is realizable by a spin space group if and only if its Brillouin-zone action contains a two-point orbit. Consequently, 842 of the 1,357 nonsymmorphic magnetic space groups are realizable by spin space groups. Applying this theory, we construct a database covering all 67,475 SSGs that can be queried bidirectionally: from an SSG to its momentum-space magnetic group, or from a given momentum-space magnetic group to all SSGs that realize it. Together, these results establish the momentum-space symmetry foundation for a systematic band theory of SSGs in spin-$1/2$ systems.
\end{abstract}

\maketitle

\textcolor{blue}{\textit{Introduction.}} 
Recently, projective crystal symmetry has attracted considerable interest in both condensed matter systems and artificial crystals~\cite{ZhaoZ2projective2020,ZhaoSwitching2021,ChenBrillouin2022,ChenClassification2023,Herzog-ArbeitmanHofstadter2023,ZhangGeneral2023,ZhangProjective2023,CalugaruMoire2025,XueProjectively2022,LiAcoustic2022,MengSpinful2023,LongNonAbelian2024,XiaoRevealing2024,HuHigherOrder2024,FonsecaWeyl2024,PuAcoustic2023,WangChessboard2023,TaoHigherorder2024,ZhuBrillouin2024,LaiRealprojectiveplane2024,LiuTopological2024,ZhangBrillouin2025,CalugaruMoire2025,DurrnagelExtended2026,DurrnagelPair2026}. A distinguishing feature of projective crystal symmetry, compared with ordinary crystal symmetry, is the emergence of nonsymmorphic symmetries in momentum space~\cite{ChenBrillouin2022}. A general theory of momentum-space nonsymmorphic groups (MNGs) within the framework of projective crystal symmetry has been established~\cite{ZhangGeneral2023}, and various MNGs have been realized in artificial crystals and condensed matter systems~\cite{XiaoSpin2024,HuHigherOrder2024,FonsecaWeyl2024,PuAcoustic2023,WangChessboard2023,TaoHigherorder2024,ZhuBrillouin2024,LaiRealprojectiveplane2024,LiuTopological2024,ZhangBrillouin2025,CalugaruMoire2025,DurrnagelExtended2026,DurrnagelPair2026}. In particular, purely nonsymmorphic groups act freely on the Brillouin torus and can therefore reduce it to other compact flat manifolds, known as platycosms~\cite{ZhangBrillouin2025}. The Klein bottle, as a representative example, has been studied extensively~\cite{ChenBrillouin2022,PuAcoustic2023,WangChessboard2023,ZhuBrillouin2024,TaoHigherorder2024}.

A common mechanism for realizing projective crystal symmetry is to introduce a gauge-flux pattern on a lattice while preserving its magnetic space group \cite{ChenClassification2023,ZhangProjective2023}. This approach generally requires fine-tuning of the hopping amplitudes. Recently, a natural mechanism for projective crystal symmetry was uncovered in the spin space groups (SSGs) of spin-$1/2$ systems \cite{XiaoSpin2024}. In this mechanism, the factor system of the magnetic space group $G$ is inherited from the intrinsic factor system associated with the double cover $SU(2)\to SO(3)$.

The concept of SSGs was introduced in the 1960s to describe magnetic systems with negligible spin-orbit coupling \cite{BrinkmanTheory1966,LitvinSpin1974,SandratskiiNoncollinear1998,LiuSpinGroup2022,SmejkalConventional2022,SmejkalEmerging2022}.  It has recently received renewed attention because of its applications to a variety of magnetic systems. In altermagnets, coplanar magnets and noncoplanar magnets, SSGs are used to describe the symmetries of magnetic orders, analyze electronic and magnon band structures \cite{CorticelliSpinspace2022,ZengDescription2024,XiaoSpin2024,ChenEnumeration2024,JiangEnumeration2024,RadaelliTensorial2024,ChenUnconventional2025,BiniskosSystematic2025}, constrain pairing channels of superconductors \cite{FengSuperconducting2025}, and analyze response functions \cite{DevarajInterplay2024,EtxebarriaCrystal2025,ElcoroAutomatic2026}. SSGs can support uncoventional spin textures, band degeneracies, quasiparticles, transport phenomena and topological states beyond those captured by magnetic space groups \cite{GuoEightfold2021,XiaoSpin2024,ChenEnumeration2024,YangSymmetry2024,LiTopological2025,YuNeel2025,TakahashiSymmetry2025,WeissenhoferMagnon2026,HuangPrediction2026,SongUnified2026}. The exhaustive enumerations of SSGs have been completed \cite{XiaoSpin2024,ChenEnumeration2024,JiangEnumeration2024}. 

Applications of SSGs to electronic structures are based on band theory and therefore require an explicit characterization of the crystalline symmetry actions on the Brillouin zone (BZ). Previous studies, however, have treated individual cases~\cite{XiaoSpin2024}, and a general theory capable of systematically deriving the MNG associated with any SSG has remained lacking. Such a theory would determine, for every SSG, the action of the magnetic point group $P$ of $G$ on the BZ. 


Here, we establish a general theory of MNGs for SSGs. We show that an MNG can arise only when the unitary translation subgroup $T_0$ of $G$ is combined with uniaxial rotations belonging to an $SO(2)$ subgroup of $SO(3)$. For every eligible SSG, we derive a compact formula [Eq.~\eqref{eq:Central-Result}] for the fractional reciprocal-lattice translation $\bm{\kappa}_p$ associated with each $p\in P$, thereby determining the $P$-action on the BZ. Our derivation reveals two structural properties. First, the factor system responsible for the nonsymmorphic action originates entirely from the double cover $\mathrm{Spin}(2)\to SO(2)$; consequently, the full structure of the double cover $SU(2)\to SO(3)$ is not required. Second, the cohomology class of $\bm{\kappa}$ is determined by whether each $p\in P$ reverses the rotation axis of the $SO(2)$ subgroup.

A natural converse question is which MNGs can be realized by SSGs---that is, what is the capacity of SSGs to realize MNGs? We answer this question with a simple theorem: an MNG is realizable by an SSG if and only if its action on the BZ admits a two-point orbit. Using our general theory, we prove this theorem and find that, among the 1,357 nonsymmorphic magnetic space groups, 842 are realizable by SSGs, whereas the remaining 515 are excluded.


\textcolor{blue}{\textit{Projective representations in SSGs}}
Spin space groups can be classified into three classes: collinear, coplanar, and noncoplanar. In a collinear SSG $\mathcal{G}$, each $g\in G$ is represented either as a purely spatial transformation or as a spatial transformation combined with the time-reversal operator $\Theta=i\sigma_2\mathcal{K}$, which reverses the spin. Such a combination does not modify the conjugation action of $G$ on the translation subgroup $T$. Consequently, the momentum-space action of $P$ is always symmorphic, as will become clear in the following section. We therefore focus hereafter on coplanar and noncoplanar SSGs.

In general, a coplanar or noncoplanar SSG $\mathcal{G}$ is isomorphic to a magnetic space group $G$~\cite{Magnetic_Grp_Notes}. 
In the present setting, $G$ is of type II when $\mathcal{G}$ is coplanar and may belong to any of the other three types when $\mathcal{G}$ is noncoplanar, in accordance with the presence or absence of time-reversal symmetry.

Each element of $\mathcal{G}$ can be written as $g\rho(g)$, where $g\in G$ is a spatial transformation and $\rho(g)\in O(3)$ acts on the internal spin degrees of freedom. The map $\rho:G\to O(3)$ preserves both group multiplication and grading: it maps unitary elements of $G$ to proper rotations and antiunitary elements to improper rotations. We therefore refer to $\rho$ as a graded homomorphism.

Whether an element $R\in O(3)$ is proper is determined by $\det R=\pm1$. Proper rotations, for which $\det R=1$, form the subgroup $SO(3)$ and act on the internal space of a spin-$1/2$ particle through a double-valued representation $D(R)\in SU(2)$. The double-valued representation is introduced in the Supplemental Material (SM)~\cite{SSGSupplement}. If $R$ is improper, it can be written as $R=(RP)P$, where $P$ denotes inversion and $RP$ is a proper rotation. We may then define $D(R)=D(RP)\Theta$. The double-valued representation $U$ of $\mathcal{G}\cong G$ on the spin-$1/2$ system is thus given by $U(g)=D[\rho(g)]$ when $g$ is unitary and by $U(g)=D[\rho(g)P]\Theta$ when $g$ is antiunitary. Because $SU(2)$ is a double cover of $SO(3)$, the representation $D$ has an intrinsic sign ambiguity. Consequently, for $g,h\in G$,
\begin{equation}
	\label{eq:factor}
	U(g)U(h)=\nu(g,h)U(gh),
\end{equation}
where $\nu(g,h)\in\{\pm1\}$. Thus, $\mathcal{G}$, or equivalently $G$, is represented projectively, with the factor system $\nu$ arising from the double cover $SU(2)\to SO(3)$~\cite{ZhangProjectiveRepresentation2026}. 

Our task is to determine the structure of the factor system $\nu$ and derive the resulting action of the magnetic point group on the BZ. In particular, we seek to identify the nonsymmorphic momentum-space actions generated by this factor system.

\textcolor{blue}{\textit{The point-group action on the Brillouin zone}}
We begin by considering the unitary subgroup $T_0$ of the translation group $T$.
Its image $\rho(T_0)$ in $SO(3)$ must be a nontrivial Abelian subgroup of
$SO(3)$. The nontrivial Abelian subgroups of $SO(3)$ can be classified into
two types~\cite{artin2011algebra}. An Abelian subgroup of the first type consists of rotations about a
common axis $\bm{n}$ and is therefore a subgroup of $SO_{\bm{n}}(2)$, the group
of all rotations about $\bm{n}$. An Abelian subgroup of the second type is
isomorphic to the Klein four-group and consists of the identity and three
$\pi$ rotations about three mutually orthogonal axes. For example,
$V_4=\{E,R_x^\pi,R_y^\pi,R_z^\pi\}$, with
$R_x^\pi R_y^\pi=R_y^\pi R_x^\pi=R_z^\pi$.

Accordingly, there are two cases to consider: either $\rho(T_0)$ is contained
in $SO_{\bm{n}}(2)$ or $\rho(T_0)$ is a Klein four-subgroup,
$\rho(T_0)\cong V_4$. In the first case, the translation operators
$U(\bm{t})$, with $\bm{t}\in T_0$, commute with one another. In the second
case, two translation operators may commute or anticommute because $V_4$ lifts
to $Q_8=\{\pm\sigma_0,\pm i\sigma_x,\pm i\sigma_y,\pm i\sigma_z\}$ in
$SU(2)$. As we show below, only the first case can give rise to MNGs.

According to Mackey's representation theory~\cite{MackeyUnitary1958,ZhangProjectiveRepresentation2026}, the BZ
is defined as the space of all irreducible representations (irreps) of $T_0$.
In the first case, the BZ is the usual torus $\widehat{T}_0$ formed by all
irreps of $T_0$. In the second case, however, the translation operators do not
necessarily commute, so the corresponding space of irreps must be constructed
differently. It suffices to consider the center $Z_{T_0}$ of the
translation-operator algebra, which consists of all translation operators that
commute with every translation operator (see the SM for more details~\cite{SSGSupplement}). The BZ is then the torus
$\widehat{Z_{T_0}}$ formed by all irreps of $Z_{T_0}$. For example, in a
nearest-neighbor tight-binding model on a rectangular lattice with a $\pi$ flux
through each plaquette, the two primitive translation operators anticommute.
The center is therefore generated by their squares, namely, by translations
through two primitive lattice spacings. 

An irrep at $\bm{k}$ in the BZ is thus given by
$e^{i\bm{k}\cdot\bm{t}}$ for $T_{\bm{t}}\in T_0$ (or
$T_{\bm{t}}\in Z_{T_0}$). We now consider the action of $P=G/T_0$ (or
$P=G/Z_{T_0}$). Here, $Z_{T_0}$ is always a normal subgroup of $G$. According
to Ref.~\cite{ZhangGeneral2023}, the action of $p\in P$ on the Brillouin torus
is determined by the conjugation of $U(\bm{t})$ by $U(\tilde{p})$, where
$\tilde{p}$ is any element of $G$ that projects to $p\in P$. This conjugation
can always be written as
\begin{equation}\label{eq:def_kappa}
	U(\tilde{p})^{-1}U(\bm{t})U(\tilde{p})
	=e^{i\bm{\kappa}_p\cdot\bm{t}}U(R_p^{-1}\bm{t}).
\end{equation}
Owing to Eq.~\eqref{eq:factor}, the left-hand side gives the phase
$\xi(\tilde{p}|\bm{t})=
\nu(\tilde{p}^{-1},T_{\bm{t}}\tilde{p})
\nu(T_{\bm{t}},\tilde{p})/\nu(\tilde{p},\tilde{p}^{-1})$.
It is straightforward to verify that
$\xi(\tilde{p}|\bm{t}_1+\bm{t}_2)=
\xi(\tilde{p}|\bm{t}_1)\xi(\tilde{p}|\bm{t}_2)$.
Moreover, $\xi(\tilde{p}|\bm{t})$ is independent of the choice of
$\tilde{p}$ because the translation operators in $T_0$ (or $Z_{T_0}$)
commute. Therefore,
$\xi(\tilde{p}|\bm{t})=e^{i\bm{\kappa}_p\cdot\bm{t}}$~\cite{ZhangGeneral2023}.
For an irrep $e^{i\bm{k}\cdot\bm{t}}$, the action of $p$ on the BZ is then
given by
\begin{equation}\label{eq:M-Action}
	p:\bm{k}\mapsto(-1)^{s_p}(R_p\bm{k}+\bm{\kappa}_p).
\end{equation}
Here, $s_p=0$ ($s_p=1$) if $p$ is unitary (antiunitary).

In the second case, the conjugation of $T_{\bm{t}}$ by $\tilde{p}$ in the
representation $U$ is always trivial because $U(\bm{t})$ is either
$\sigma_0$ or $-\sigma_0$ for every $\bm{t}\in Z_{T_0}$. Consequently, the
$P$-action on the BZ cannot be nonsymmorphic. Henceforth, we focus on the first
case, in which the unitary component $T_0$ of the translation subgroup is
mapped to rotations in $SO_{\bm{n}}(2)$ about a common axis $\bm{n}$.

\textcolor{blue}{\textit{Fractional reciprocal-lattice translations}}
We now derive an explicit formula for $\bm{\kappa}_p$. The restriction of the
homomorphism $\rho$ to $T_0$ can be written as
\begin{equation}\label{eq:Tn}
	\rho(\bm{t})=e^{-i\bm{\beta}\cdot\bm{t}\hat{J}_{\bm{n}}}.
\end{equation}
Here, $\hat{J}_{\bm{n}}$ is the Lie-algebra generator of
$SO_{\bm{n}}(2)$, and $\theta(\bm{t})=\bm{\beta}\cdot\bm{t}$ is the
rotation angle associated with the translation $\bm{t}$, with $\bm{\beta}$ a
vector in the Brillouin torus. This form of $\theta(\bm{t})$ follows from the
homomorphism condition
$\theta(\bm{t}_1+\bm{t}_2)=\theta(\bm{t}_1)+\theta(\bm{t}_2)\mod 2\pi$.

The image $\rho(G)$ of $G$ in $O(3)$ can be characterized elegantly. Because
$T_0$ is a normal subgroup of $G$, $\rho(T_0)$ must also be a normal subgroup
of $\rho(G)$. Therefore, $\rho(G)$ is contained in the normalizer
$N(SO_{\bm{n}}(2))$ of $SO_{\bm{n}}(2)$ in $O(3)$, which consists of all
elements of $O(3)$ whose conjugation leaves $SO_{\bm{n}}(2)$ invariant. The
normalizer has a simple structure: in addition to $SO_{\bm{n}}(2)$ and
inversion $P$, only a $\pi$ rotation $R_{\bm{m}}^\pi$ about any axis
$\bm{m}$ orthogonal to $\bm{n}$ is required to generate the entire group~\cite{artin2011algebra},
namely,
\begin{equation}
	N(SO_{\bm{n}}(2))
	=\langle SO_{\bm{n}}(2),R_{\bm{m}}(\pi),P\rangle.
	\label{eq:normalizer}
\end{equation}
Because
$R_{\bm{m}}^\pi\hat{J}_{\bm{n}}R_{\bm{m}}^\pi=-\hat{J}_{\bm{n}}$,
conjugation by $R_{\bm{m}}^\pi$ reverses the rotation angle about $\bm{n}$.
The elements of $G$, or equivalently of $P\cong G/T_0$, can therefore be
graded according to whether $\rho(g)$ or $\rho(\tilde{p})$ reverses this
angle:
\begin{equation}
	\rho(\tilde{p})^{-1}
	e^{-i\bm{\beta}\cdot\bm{t}\hat{J}_{\bm{n}}}
	\rho(\tilde{p})
	=e^{-i\eta_p\bm{\beta}\cdot\bm{t}\hat{J}_{\bm{n}}},
\end{equation}
where $\eta$ is a homomorphism from $P\cong G/T_0$ to
$C_2=\{\pm1\}$. 
On the other hand, the group relation
$\tilde{p}^{-1}T_{\bm{t}}\tilde{p}=T_{R_p^{-1}\bm{t}}$ implies
\begin{equation}\label{1}
	\rho(\tilde{p})^{-1}
	e^{-i\bm{\beta}\cdot\bm{t}\hat{J}_{\bm{n}}}
	\rho(\tilde{p})
	=e^{-i\bm{\beta}\cdot R_p^{-1}\bm{t}\hat{J}_{\bm{n}}}
	=e^{-i(R_p\bm{\beta})\cdot\bm{t}\hat{J}_{\bm{n}}}.
\end{equation}
Together, these two identities imply
$(\eta_p\bm{\beta}-R_p\bm{\beta})\cdot\bm{t}=0\mod 2\pi$.
Thus, $\eta_p\bm{\beta}-R_p\bm{\beta}$ belongs to the reciprocal lattice
$L_F$:
\begin{equation}
	\eta_p\bm{\beta}-R_p\bm{\beta}\in L_F.
	\label{eq:quantization}
\end{equation}

We can now derive $\bm{\kappa}_p$ by lifting the conjugation to the
representation $U$. The fractional translations $\bm{\kappa}_p$ originate
entirely from the double cover of $SO_{\bm{n}}(2)$ by
$\mathrm{Spin}_{\bm{n}}(2)$ because the sign ambiguity in $U(\tilde{p})$
cancels in the conjugation. With
$\sigma_{\bm{n}}/2=\bm{n}\cdot\bm{\sigma}/2$, we have
\begin{equation}
	U(\bm{t})=e^{-i\bm{\beta}\cdot\bm{t}\sigma_{\bm{n}}/2},
\end{equation}
and therefore
\begin{equation}
	U(\tilde{p})^{-1}
	e^{-i\bm{\beta}\cdot\bm{t}\sigma_{\bm{n}}/2}
	U(\tilde{p})
	=e^{-i\eta_p\bm{\beta}\cdot\bm{t}\sigma_{\bm{n}}/2}.
\end{equation}
This relation can be written as
\begin{equation}
	U(\tilde{p})^{-1}U(\bm{t})U(\tilde{p})
	=e^{-i\bm{\kappa}_p\cdot\bm{t}\sigma_{\bm{n}}}
	U(R_p^{-1}\bm{t})
	\label{eq:xi}
\end{equation}
with
\begin{equation}\label{eq:Central-Result}
	\bm{\kappa}_p
	=\frac{1}{2}(\eta_p\bm{\beta}-R_p\bm{\beta})\mod L_F.
\end{equation}
By Eq.~\eqref{eq:quantization}, $\bm{\kappa}_p\cdot\bm{t}$ is an integer
multiple of $\pi$, and hence
$e^{-i\bm{\kappa}_p\cdot\bm{t}\sigma_{\bm{n}}}
=e^{-i\bm{\kappa}_p\cdot\bm{t}}=\pm1$.
In particular, $\bm{\kappa}_p$ takes values in $L_F/2$. We have thus
established Eq.~\eqref{eq:def_kappa}; the $P$-action on the BZ in
Eq.~\eqref{eq:M-Action} is therefore completely specified by
Eq.~\eqref{eq:Central-Result}.

Moreover, using $\eta_{p_1p_2}=\eta_{p_1}\eta_{p_2}$ and
$R_{p_1p_2}=R_{p_1}R_{p_2}$, we find that $\bm{\kappa}_p$ satisfies
\begin{equation}\label{eq:cocycle-derivation}
	R_{p_1}\bm{\kappa}_{p_2}
	-\bm{\kappa}_{p_1p_2}+\bm{\kappa}_{p_1}
	=(1-\eta_{p_2})\bm{\kappa}_{p_1}
	\in L_F,
\end{equation}
which ensures that the $P$-action in Eq.~\eqref{eq:M-Action} is well defined
on the BZ torus. 

\begin{table}
	\caption{Realizability of MNGs of the four types. The first column specifies
		the group type, the second gives the number of nonsymmorphic groups of
		each type, and the third and fourth give the numbers of realizable and
		nonrealizable groups, respectively.}
	\label{tab:orbit2}
	\begin{ruledtabular}
		\begin{tabular}{c@{\quad}r@{\quad}r@{\quad}r}
			Type & Total & Realizable & Nonrealizable \\
			\hline
			I   & 157 & 109 & 48 \\
			II  & 157 & 109 & 48 \\
			III & 526 & 386 & 140 \\
			IV  & 517 & 238 & 279 \\
			\hline
			All & 1357 & 842 & 515
		\end{tabular}
	\end{ruledtabular}
\end{table}

\textcolor{blue}{\textit{Realizability of MNGs}}
Extending the $P$-action of an SSG on the BZ by the reciprocal-lattice
translations $T_F$ yields a momentum-space magnetic space group (MG). Not
every MNG can arise from an SSG, so a natural question is which MNGs can be
realized by SSGs. The following theorem provides a simple answer.

\textbf{Two-point-orbit theorem.}\quad
\textit{A momentum-space nonsymmorphic magnetic space group can be realized by
	a spin space group if and only if its action on the Brillouin zone has a
	two-point orbit.}

We first show that every realizable MNG, whether magnetic or nonmagnetic, has a
two-point orbit in the BZ under the $P$-action. To this end, we regard
$\bm{k}_0=\bm{\beta}/2$ as a wave vector in the BZ and consider its orbit.
Equation~\eqref{eq:Central-Result} implies that
\begin{equation}
	R_p\bm{k}_0+\bm{\kappa}_p
	=\eta_p\bm{k}_0\mod L_F
\end{equation}
for every $p\in P$. Including the factor $(-1)^{s_p}$ in
Eq.~\eqref{eq:M-Action}, $p$ maps $\bm{k}_0$ to
$(-1)^{s_p}\eta_p\bm{k}_0$. Thus, the orbit of $\bm{k}_0$ under the
$P$-action consists of the two points $\bm{k}_0$ and $-\bm{k}_0$, because the
grading $\eta$ of $P$ is nontrivial for an MNG.

Conversely, suppose that an MNG has a two-point orbit
$\{\bm{k}_+,\bm{k}_-\}$. We construct an SSG that realizes this MNG. The
magnetic space group $G$ of the SSG is chosen to be the symmorphic group in the
arithmetic class dual to that of the MNG. It then remains to construct a map
$\rho$ that maps $T_0$ and $P$ appropriately to $O(3)$ so as to realize the
MNG.

\begin{table}
	\caption{Momentum-space purely nonsymmorphic groups and their SSG
		realizations. The first column lists the corresponding Brillouin
		platycosms, the second gives the BNS symbol and group type, and the third
		lists all SSGs that realize each MNG, using the labels of
		Ref.~\cite{XiaoSpin2024}.}
	\label{tab:platycosms}
	\scriptsize
	\setlength{\tabcolsep}{3pt}
	\begin{ruledtabular}
		\begin{tabular}{p{0.16\columnwidth} l p{0.47\columnwidth}}
			Platycosm & BNS (Type) & SSG Labels \\
			\hline
			Torocosm & $\mathrm{P}1$ 1.1 (I) & L1.1.1, N1.1.1, N1.3.1, N1.5.1, N1.9.4 \\
			& $\mathrm{P}_{S}1$ 1.3 (IV) & N2.9.3 \\
			Dicosm & $\mathrm{P}2_1$ 4.7 (I) & N3.9.15, N3.9.20, N3.14.2 \\
			& $\mathrm{P}2_1'$ 4.9 (III) & N6.9.18, N6.9.24, N6.11.3, N7.9.5, N7.9.12, N7.10.2,
			N7.11.2 \\
			\multirow{2}{*}{\shortstack[l]{First\\amphicosm}} & $\mathrm{Pc}$ 7.24 (I) & N6.9.8, N6.9.19, N6.14.2, N7.9.6,
			N7.9.10, N7.14.2 \\
			& $\mathrm{Pc}'$ 7.26 (III) & N3.9.6, N3.9.14, N3.11.3, N4.9.3, N4.10.2, N4.11.2 \\
			\multirow{2}{*}{\shortstack[l]{Second\\amphicosm}} & $\mathrm{Cc}$ 9.37 (I) & N8.9.8, N8.14.2 \\
			& $\mathrm{Cc}'$ 9.39 (III) & N5.9.5, N5.11.3
		\end{tabular}
	\end{ruledtabular}
\end{table}

In this construction, we choose the midpoint of the line segment joining
$\bm{k}_+$ and $\bm{k}_-$ as the origin of momentum space. Whether $p\in P$ exchanges the two points then defines a natural grading of
$P$, corresponding to $(-1)^{s_p}\eta_p$. Hence,
$(-1)^{s_p}(R_p\bm{k}_++\bm{\kappa}_p)=(-1)^{s_p}\eta_p\bm{k}_+$ leads to
\begin{equation}
	\bm{\kappa}_p
	=(\eta_p-R_p)\bm{k}_+.
\end{equation}
Introducing $\bm{\beta}=2\bm{k}_+$, we define $\rho$ by
\begin{equation}
	\rho(\bm{t})=e^{-i(\bm{\beta}\cdot\bm{t})\hat{J}_z},\qquad
	\rho(p)=\operatorname{diag}
	(1,\eta_p,(-1)^{s_p}\eta_p).
	\label{eq:orbit-construction}
\end{equation}
It is straightforward to verify that an SSG with the symmorphic group $G$ and
the homomorphism $\rho$ realizes the given MNG with the two-point
orbit. 

This proves the two-point-orbit theorem. Applying the theorem to all 1,357
MNGs, we find that 842 have a two-point orbit and are therefore realizable by
SSGs, whereas the remaining 515 have no such orbit and are not realizable. The
distribution among the four types of magnetic space groups is given in
Table~\ref{tab:orbit2}. The group-by-group results are listed in SM~\cite{SSGSupplement}.

\textcolor{blue}{\textit{Examples.---}}
We now illustrate our theory through a systematic study of momentum-space
purely nonsymmorphic groups. The orbit spaces of these groups are compact flat
three-manifolds known as platycosms~\cite{ZhangBrillouin2025}. Type-I purely
nonsymmorphic groups have been studied in general; here, we investigate all
types realizable by SSGs.

By the two-point-orbit theorem, only four of the nine possible purely
nonsymmorphic actions, which together with $P1$ correspond to the ten
Bieberbach groups, are realizable by SSGs. Each of these four actions
corresponds to a platycosm. Collectively, the four actions correspond to eight
magnetic space groups, each of which can be realized by several SSGs. In total,
32 SSGs realize momentum-space purely nonsymmorphic groups. These results are
listed in Table~\ref{tab:platycosms}. Below, we present two examples of the two
MNGs associated with the dicosm; the
remaining cases are discussed in the SM~\cite{SSGSupplement}.

The first example is SSG N3.9.15, with $G=P2$ (BNS 3.1, type I). The group
$P2$ is symmorphic, with point group $P=C_{2y}$. The group $C_{2y}$ is
generated by the $\pi$ rotation $p$ about the $y$ axis, for which
$R_p=\operatorname{diag}(-1,1,-1)$. In SSG N3.9.15, all translations in $T$
are combined with spin rotations about the $z$ axis, with
$\bm{\beta}=(0,\pi,0)$, while $p$ is combined with the spin rotation
$\rho(p)=R_y^\pi$, yielding $\eta_p=-1$. Equation~\eqref{eq:Central-Result}
then gives
\begin{equation}\label{eq:kappa_p}
	\bm{\kappa}_p
	=\frac{1}{2}(\eta_p\bm{\beta}-R_p\bm{\beta})
	=(0,\pi,0),
\end{equation}
which corresponds to a twofold screw rotation along the $k_y$ axis:
\begin{equation}\label{eq:screw}
	p:(k_x,k_y,k_z)\mapsto(-k_x,k_y+\pi,-k_z).
\end{equation}
Here, all real-space lattice constants are set to $1$, so the reciprocal-lattice
constants are $2\pi$. The resulting MNG is $P2_1$
(BNS 4.7, type I), which corresponds to the dicosm.

\begin{table}
	\caption{MGs realized by all SSGs, classified into four types. The first
		column specifies the type of the realized MGs, the second gives the total
		number of SSGs realizing MGs of each type, and the third and fourth give
		the numbers with symmorphic and nonsymmorphic momentum-space groups,
		respectively.}
	\label{tab:ssg_census}
	\begin{ruledtabular}
		\begin{tabular}{c r r r}
			MG Type & Total & Symmorphic & Nonsymmorphic \\
			\hline
			I   & 5535  & 4441  & 1094 \\
			II  & 27647 & 23841 & 3806 \\
			III & 27549 & 20392 & 7157 \\
			IV  & 6744  & 0     & 6744 \\
			\hline
			All & 67475 & 48674 & 18801
		\end{tabular}
	\end{ruledtabular}
\end{table}

The second example is SSG N6.9.18, with $G=Pm'$ (BNS 6.20, type III). Its
magnetic point group $P$ is generated by an antiunitary twofold element $p$
with $R_p=\operatorname{diag}(1,-1,1)$ and $s_p=1$; spatially, it acts as a
mirror reflection that reverses the $y$ coordinate. The element $p$ is
combined with the improper rotation
$\rho(p)=\operatorname{diag}(-1,1,1)$. The translations are combined with
rotations about the $x$ axis, with $\bm{\beta}=(\pi,\pi,0)$. Because
$\rho(p)$ preserves the rotation axis, $\eta_p=1$. Equation~\eqref{eq:Central-Result}
then gives the same result as Eq.~\eqref{eq:kappa_p}. Because $p$ is
antiunitary, Eq.~\eqref{eq:M-Action} maps $\bm{k}$ to
$-R_p\bm{k}-\bm{\kappa}_p$. The action of $p$ is therefore again the twofold
screw rotation in Eq.~\eqref{eq:screw}, but it now acts antiunitarily on the
BZ.

\textcolor{blue}{\textit{Discussion and conclusion.}}
In summary, we have established a general theory of momentum-space symmetry for coplanar and noncoplanar SSGs. We showed that an SSG can give rise to an MNG only when its unitary translations are combined with rotations about a common axis. We further demonstrated that the nonsymmorphic action originates entirely from the double cover $\mathrm{Spin}(2)\to SO(2)$, and derived a general formula for all such nonsymmorphic actions on the BZ. The theory also yields the two-point-orbit theorem, which provides an immediate criterion for determining whether a given MNG can be realized by an SSG.

Having classified the nontrivial MNGs, we can determine the complete momentum-space magnetic group (MG) associated with every SSG. We have constructed a reciprocal-space SSG database covering all 67,475 SSGs enumerated in Ref.~\cite{XiaoSpin2024}. For each SSG, the database provides the unitary central lattice, the construction of the BZ, the $P$-action on the BZ, and the resulting momentum-space MG. The database can be queried in both directions: from an SSG to its corresponding MG, or from a given MG to all SSGs that realize it. The resulting statistics are summarized in Table~\ref{tab:ssg_census}. Among the 67,475 SSGs, 18,801 exhibit MNGs, whereas 48,674 have symmorphic momentum-space magnetic groups. Altogether, SSGs realize 1,136 distinct MGs, which fall into four types.

Our unified theory and the resulting database of momentum-space MGs provide a symmetry foundation for studying the band structures of spin-$1/2$ systems with arbitrary SSGs. They are therefore applicable to a broad range of problems involving SSG-based band theory. Together with the representations of the little cogroups on the BZ, the momentum-space MG associated with an SSG can be used to construct $k\cdot p$ models at high-symmetry momenta, characterize emergent quasiparticles, and diagnose crystalline topological phases~\cite{ZhangProjectiveRepresentation2026}.


\bibliography{references}

\begin{thebibliography}{59}%
\makeatletter
\providecommand \@ifxundefined [1]{%
 \@ifx{#1\undefined}
}%
\providecommand \@ifnum [1]{%
 \ifnum #1\expandafter \@firstoftwo
 \else \expandafter \@secondoftwo
 \fi
}%
\providecommand \@ifx [1]{%
 \ifx #1\expandafter \@firstoftwo
 \else \expandafter \@secondoftwo
 \fi
}%
\providecommand \natexlab [1]{#1}%
\providecommand \enquote  [1]{``#1''}%
\providecommand \bibnamefont  [1]{#1}%
\providecommand \bibfnamefont [1]{#1}%
\providecommand \citenamefont [1]{#1}%
\providecommand \href@noop [0]{\@secondoftwo}%
\providecommand \href [0]{\begingroup \@sanitize@url \@href}%
\providecommand \@href[1]{\@@startlink{#1}\@@href}%
\providecommand \@@href[1]{\endgroup#1\@@endlink}%
\providecommand \@sanitize@url [0]{\catcode `\\12\catcode `\$12\catcode `\&12\catcode `\#12\catcode `\^12\catcode `\_12\catcode `\%12\relax}%
\providecommand \@@startlink[1]{}%
\providecommand \@@endlink[0]{}%
\providecommand \url  [0]{\begingroup\@sanitize@url \@url }%
\providecommand \@url [1]{\endgroup\@href {#1}{\urlprefix }}%
\providecommand \urlprefix  [0]{URL }%
\providecommand \Eprint [0]{\href }%
\providecommand \doibase [0]{https://doi.org/}%
\providecommand \selectlanguage [0]{\@gobble}%
\providecommand \bibinfo  [0]{\@secondoftwo}%
\providecommand \bibfield  [0]{\@secondoftwo}%
\providecommand \translation [1]{[#1]}%
\providecommand \BibitemOpen [0]{}%
\providecommand \bibitemStop [0]{}%
\providecommand \bibitemNoStop [0]{.\EOS\space}%
\providecommand \EOS [0]{\spacefactor3000\relax}%
\providecommand \BibitemShut  [1]{\csname bibitem#1\endcsname}%
\let\auto@bib@innerbib\@empty
\bibitem [{\citenamefont {Zhao}\ \emph {et~al.}(2020)\citenamefont {Zhao}, \citenamefont {Huang},\ and\ \citenamefont {Yang}}]{ZhaoZ2projective2020}%
  \BibitemOpen
  \bibfield  {author} {\bibinfo {author} {\bibfnamefont {Y.~X.}\ \bibnamefont {Zhao}}, \bibinfo {author} {\bibfnamefont {Y.-X.}\ \bibnamefont {Huang}},\ and\ \bibinfo {author} {\bibfnamefont {S.~A.}\ \bibnamefont {Yang}},\ }\bibfield  {title} {\bibinfo {title} {Z2-projective translational symmetry protected topological phases},\ }\href {https://doi.org/10.1103/PhysRevB.102.161117} {\bibfield  {journal} {\bibinfo  {journal} {Phys. Rev. B}\ }\textbf {\bibinfo {volume} {102}},\ \bibinfo {pages} {161117} (\bibinfo {year} {2020})}\BibitemShut {NoStop}%
\bibitem [{\citenamefont {Zhao}\ \emph {et~al.}(2021)\citenamefont {Zhao}, \citenamefont {Chen}, \citenamefont {Sheng},\ and\ \citenamefont {Yang}}]{ZhaoSwitching2021}%
  \BibitemOpen
  \bibfield  {author} {\bibinfo {author} {\bibfnamefont {Y.~X.}\ \bibnamefont {Zhao}}, \bibinfo {author} {\bibfnamefont {C.}~\bibnamefont {Chen}}, \bibinfo {author} {\bibfnamefont {X.-L.}\ \bibnamefont {Sheng}},\ and\ \bibinfo {author} {\bibfnamefont {S.~A.}\ \bibnamefont {Yang}},\ }\bibfield  {title} {\bibinfo {title} {Switching {{Spinless}} and {{Spinful Topological Phases}} with {{Projective PT Symmetry}}},\ }\href {https://doi.org/10.1103/PhysRevLett.126.196402} {\bibfield  {journal} {\bibinfo  {journal} {Phys. Rev. Lett.}\ }\textbf {\bibinfo {volume} {126}},\ \bibinfo {pages} {196402} (\bibinfo {year} {2021})}\BibitemShut {NoStop}%
\bibitem [{\citenamefont {Chen}\ \emph {et~al.}(2022)\citenamefont {Chen}, \citenamefont {Yang},\ and\ \citenamefont {Zhao}}]{ChenBrillouin2022}%
  \BibitemOpen
  \bibfield  {author} {\bibinfo {author} {\bibfnamefont {Z.~Y.}\ \bibnamefont {Chen}}, \bibinfo {author} {\bibfnamefont {S.~A.}\ \bibnamefont {Yang}},\ and\ \bibinfo {author} {\bibfnamefont {Y.~X.}\ \bibnamefont {Zhao}},\ }\bibfield  {title} {\bibinfo {title} {Brillouin {{Klein}} bottle from artificial gauge fields},\ }\href {https://doi.org/10.1038/s41467-022-29953-7} {\bibfield  {journal} {\bibinfo  {journal} {Nat. Commun.}\ }\textbf {\bibinfo {volume} {13}},\ \bibinfo {pages} {2215} (\bibinfo {year} {2022})}\BibitemShut {NoStop}%
\bibitem [{\citenamefont {Chen}\ \emph {et~al.}(2023)\citenamefont {Chen}, \citenamefont {Zhang}, \citenamefont {Yang},\ and\ \citenamefont {Zhao}}]{ChenClassification2023}%
  \BibitemOpen
  \bibfield  {author} {\bibinfo {author} {\bibfnamefont {Z.~Y.}\ \bibnamefont {Chen}}, \bibinfo {author} {\bibfnamefont {Z.}~\bibnamefont {Zhang}}, \bibinfo {author} {\bibfnamefont {S.~A.}\ \bibnamefont {Yang}},\ and\ \bibinfo {author} {\bibfnamefont {Y.~X.}\ \bibnamefont {Zhao}},\ }\bibfield  {title} {\bibinfo {title} {Classification of time-reversal-invariant crystals with gauge structures},\ }\href {https://doi.org/10.1038/s41467-023-36447-7} {\bibfield  {journal} {\bibinfo  {journal} {Nat. Commun.}\ }\textbf {\bibinfo {volume} {14}},\ \bibinfo {pages} {743} (\bibinfo {year} {2023})}\BibitemShut {NoStop}%
\bibitem [{\citenamefont {{Herzog-Arbeitman}}\ \emph {et~al.}(2023)\citenamefont {{Herzog-Arbeitman}}, \citenamefont {Song}, \citenamefont {Elcoro},\ and\ \citenamefont {Bernevig}}]{Herzog-ArbeitmanHofstadter2023}%
  \BibitemOpen
  \bibfield  {author} {\bibinfo {author} {\bibfnamefont {J.}~\bibnamefont {{Herzog-Arbeitman}}}, \bibinfo {author} {\bibfnamefont {Z.-D.}\ \bibnamefont {Song}}, \bibinfo {author} {\bibfnamefont {L.}~\bibnamefont {Elcoro}},\ and\ \bibinfo {author} {\bibfnamefont {B.~A.}\ \bibnamefont {Bernevig}},\ }\bibfield  {title} {\bibinfo {title} {Hofstadter {{Topology}} with {{Real Space Invariants}} and {{Reentrant Projective Symmetries}}},\ }\href {https://doi.org/10.1103/PhysRevLett.130.236601} {\bibfield  {journal} {\bibinfo  {journal} {Phys. Rev. Lett.}\ }\textbf {\bibinfo {volume} {130}},\ \bibinfo {pages} {236601} (\bibinfo {year} {2023})}\BibitemShut {NoStop}%
\bibitem [{\citenamefont {Zhang}\ \emph {et~al.}(2023{\natexlab{a}})\citenamefont {Zhang}, \citenamefont {Chen}, \citenamefont {Zhang},\ and\ \citenamefont {Zhao}}]{ZhangGeneral2023}%
  \BibitemOpen
  \bibfield  {author} {\bibinfo {author} {\bibfnamefont {C.}~\bibnamefont {Zhang}}, \bibinfo {author} {\bibfnamefont {Z.~Y.}\ \bibnamefont {Chen}}, \bibinfo {author} {\bibfnamefont {Z.}~\bibnamefont {Zhang}},\ and\ \bibinfo {author} {\bibfnamefont {Y.~X.}\ \bibnamefont {Zhao}},\ }\bibfield  {title} {\bibinfo {title} {General {{Theory}} of {{Momentum-Space Nonsymmorphic Symmetry}}},\ }\href {https://doi.org/10.1103/PhysRevLett.130.256601} {\bibfield  {journal} {\bibinfo  {journal} {Phys. Rev. Lett.}\ }\textbf {\bibinfo {volume} {130}},\ \bibinfo {pages} {256601} (\bibinfo {year} {2023}{\natexlab{a}})}\BibitemShut {NoStop}%
\bibitem [{\citenamefont {Zhang}\ \emph {et~al.}(2023{\natexlab{b}})\citenamefont {Zhang}, \citenamefont {Chen},\ and\ \citenamefont {Zhao}}]{ZhangProjective2023}%
  \BibitemOpen
  \bibfield  {author} {\bibinfo {author} {\bibfnamefont {Z.}~\bibnamefont {Zhang}}, \bibinfo {author} {\bibfnamefont {Z.~Y.}\ \bibnamefont {Chen}},\ and\ \bibinfo {author} {\bibfnamefont {Y.~X.}\ \bibnamefont {Zhao}},\ }\bibfield  {title} {\bibinfo {title} {Projective spacetime symmetry of spacetime crystals},\ }\href {https://doi.org/10.1038/s42005-023-01446-z} {\bibfield  {journal} {\bibinfo  {journal} {Commun. Phys.}\ }\textbf {\bibinfo {volume} {6}},\ \bibinfo {pages} {326} (\bibinfo {year} {2023}{\natexlab{b}})}\BibitemShut {NoStop}%
\bibitem [{\citenamefont {C{\u a}lug{\u a}ru}\ \emph {et~al.}(2025)\citenamefont {C{\u a}lug{\u a}ru}, \citenamefont {Jiang}, \citenamefont {Hu}, \citenamefont {Pi}, \citenamefont {Yu}, \citenamefont {Vergniory}, \citenamefont {Shan}, \citenamefont {Felser}, \citenamefont {Schoop}, \citenamefont {Efetov}, \citenamefont {Mak},\ and\ \citenamefont {Bernevig}}]{CalugaruMoire2025}%
  \BibitemOpen
  \bibfield  {author} {\bibinfo {author} {\bibfnamefont {D.}~\bibnamefont {C{\u a}lug{\u a}ru}}, \bibinfo {author} {\bibfnamefont {Y.}~\bibnamefont {Jiang}}, \bibinfo {author} {\bibfnamefont {H.}~\bibnamefont {Hu}}, \bibinfo {author} {\bibfnamefont {H.}~\bibnamefont {Pi}}, \bibinfo {author} {\bibfnamefont {J.}~\bibnamefont {Yu}}, \bibinfo {author} {\bibfnamefont {M.~G.}\ \bibnamefont {Vergniory}}, \bibinfo {author} {\bibfnamefont {J.}~\bibnamefont {Shan}}, \bibinfo {author} {\bibfnamefont {C.}~\bibnamefont {Felser}}, \bibinfo {author} {\bibfnamefont {L.~M.}\ \bibnamefont {Schoop}}, \bibinfo {author} {\bibfnamefont {D.~K.}\ \bibnamefont {Efetov}}, \bibinfo {author} {\bibfnamefont {K.~F.}\ \bibnamefont {Mak}},\ and\ \bibinfo {author} {\bibfnamefont {B.~A.}\ \bibnamefont {Bernevig}},\ }\bibfield  {title} {\bibinfo {title} {Moir\'e materials based on {{M-point}} twisting},\ }\href {https://doi.org/10.1038/s41586-025-09187-5} {\bibfield  {journal} {\bibinfo  {journal} {Nature}\ }\textbf {\bibinfo {volume} {643}},\ \bibinfo {pages} {376} (\bibinfo {year} {2025})}\BibitemShut {NoStop}%
\bibitem [{\citenamefont {Xue}\ \emph {et~al.}(2022)\citenamefont {Xue}, \citenamefont {Wang}, \citenamefont {Huang}, \citenamefont {Cheng}, \citenamefont {Yu}, \citenamefont {Foo}, \citenamefont {Zhao}, \citenamefont {Yang},\ and\ \citenamefont {Zhang}}]{XueProjectively2022}%
  \BibitemOpen
  \bibfield  {author} {\bibinfo {author} {\bibfnamefont {H.}~\bibnamefont {Xue}}, \bibinfo {author} {\bibfnamefont {Z.}~\bibnamefont {Wang}}, \bibinfo {author} {\bibfnamefont {Y.-X.}\ \bibnamefont {Huang}}, \bibinfo {author} {\bibfnamefont {Z.}~\bibnamefont {Cheng}}, \bibinfo {author} {\bibfnamefont {L.}~\bibnamefont {Yu}}, \bibinfo {author} {\bibfnamefont {Y.~X.}\ \bibnamefont {Foo}}, \bibinfo {author} {\bibfnamefont {Y.~X.}\ \bibnamefont {Zhao}}, \bibinfo {author} {\bibfnamefont {S.~A.}\ \bibnamefont {Yang}},\ and\ \bibinfo {author} {\bibfnamefont {B.}~\bibnamefont {Zhang}},\ }\bibfield  {title} {\bibinfo {title} {Projectively {{Enriched Symmetry}} and {{Topology}} in {{Acoustic Crystals}}},\ }\href {https://doi.org/10.1103/PhysRevLett.128.116802} {\bibfield  {journal} {\bibinfo  {journal} {Phys. Rev. Lett.}\ }\textbf {\bibinfo {volume} {128}},\ \bibinfo {pages} {116802} (\bibinfo {year} {2022})}\BibitemShut {NoStop}%
\bibitem [{\citenamefont {Li}\ \emph {et~al.}(2022)\citenamefont {Li}, \citenamefont {Du}, \citenamefont {Zhang}, \citenamefont {Li}, \citenamefont {Fan}, \citenamefont {Zhang},\ and\ \citenamefont {Qiu}}]{LiAcoustic2022}%
  \BibitemOpen
  \bibfield  {author} {\bibinfo {author} {\bibfnamefont {T.}~\bibnamefont {Li}}, \bibinfo {author} {\bibfnamefont {J.}~\bibnamefont {Du}}, \bibinfo {author} {\bibfnamefont {Q.}~\bibnamefont {Zhang}}, \bibinfo {author} {\bibfnamefont {Y.}~\bibnamefont {Li}}, \bibinfo {author} {\bibfnamefont {X.}~\bibnamefont {Fan}}, \bibinfo {author} {\bibfnamefont {F.}~\bibnamefont {Zhang}},\ and\ \bibinfo {author} {\bibfnamefont {C.}~\bibnamefont {Qiu}},\ }\bibfield  {title} {\bibinfo {title} {Acoustic {{M\"obius}} insulators from projective symmetry},\ }\href {https://doi.org/10.1103/PhysRevLett.128.116803} {\bibfield  {journal} {\bibinfo  {journal} {Phys. Rev. Lett.}\ }\textbf {\bibinfo {volume} {128}},\ \bibinfo {pages} {116803} (\bibinfo {year} {2022})}\BibitemShut {NoStop}%
\bibitem [{\citenamefont {Meng}\ \emph {et~al.}(2023)\citenamefont {Meng}, \citenamefont {Lin}, \citenamefont {Shi}, \citenamefont {Wei}, \citenamefont {Yang}, \citenamefont {Yan}, \citenamefont {Zhu}, \citenamefont {Xi}, \citenamefont {Wang}, \citenamefont {Ge}, \citenamefont {Yuan}, \citenamefont {Chen}, \citenamefont {Liu}, \citenamefont {Sun}, \citenamefont {Chen}, \citenamefont {Yang},\ and\ \citenamefont {Gao}}]{MengSpinful2023}%
  \BibitemOpen
  \bibfield  {author} {\bibinfo {author} {\bibfnamefont {Y.}~\bibnamefont {Meng}}, \bibinfo {author} {\bibfnamefont {S.}~\bibnamefont {Lin}}, \bibinfo {author} {\bibfnamefont {B.-j.}\ \bibnamefont {Shi}}, \bibinfo {author} {\bibfnamefont {B.}~\bibnamefont {Wei}}, \bibinfo {author} {\bibfnamefont {L.}~\bibnamefont {Yang}}, \bibinfo {author} {\bibfnamefont {B.}~\bibnamefont {Yan}}, \bibinfo {author} {\bibfnamefont {Z.}~\bibnamefont {Zhu}}, \bibinfo {author} {\bibfnamefont {X.}~\bibnamefont {Xi}}, \bibinfo {author} {\bibfnamefont {Y.}~\bibnamefont {Wang}}, \bibinfo {author} {\bibfnamefont {Y.}~\bibnamefont {Ge}}, \bibinfo {author} {\bibfnamefont {S.-q.}\ \bibnamefont {Yuan}}, \bibinfo {author} {\bibfnamefont {J.}~\bibnamefont {Chen}}, \bibinfo {author} {\bibfnamefont {G.-G.}\ \bibnamefont {Liu}}, \bibinfo {author} {\bibfnamefont {H.-x.}\ \bibnamefont {Sun}}, \bibinfo {author} {\bibfnamefont {H.}~\bibnamefont {Chen}}, \bibinfo {author} {\bibfnamefont {Y.}~\bibnamefont {Yang}},\ and\ \bibinfo {author} {\bibfnamefont {Z.}~\bibnamefont {Gao}},\ }\bibfield  {title} {\bibinfo {title} {Spinful {{Topological Phases}} in {{Acoustic Crystals}} with {{Projective PT Symmetry}}},\ }\href {https://doi.org/10.1103/PhysRevLett.130.026101} {\bibfield  {journal} {\bibinfo  {journal} {Phys. Rev. Lett.}\ }\textbf {\bibinfo {volume} {130}},\ \bibinfo {pages} {026101} (\bibinfo {year} {2023})}\BibitemShut {NoStop}%
\bibitem [{\citenamefont {Long}\ \emph {et~al.}(2024)\citenamefont {Long}, \citenamefont {Wang}, \citenamefont {Zhang}, \citenamefont {Xue}, \citenamefont {Zhao},\ and\ \citenamefont {Zhang}}]{LongNonAbelian2024}%
  \BibitemOpen
  \bibfield  {author} {\bibinfo {author} {\bibfnamefont {Y.}~\bibnamefont {Long}}, \bibinfo {author} {\bibfnamefont {Z.}~\bibnamefont {Wang}}, \bibinfo {author} {\bibfnamefont {C.}~\bibnamefont {Zhang}}, \bibinfo {author} {\bibfnamefont {H.}~\bibnamefont {Xue}}, \bibinfo {author} {\bibfnamefont {Y.~X.}\ \bibnamefont {Zhao}},\ and\ \bibinfo {author} {\bibfnamefont {B.}~\bibnamefont {Zhang}},\ }\bibfield  {title} {\bibinfo {title} {Non-{{Abelian Braiding}} of {{Topological Edge Bands}}},\ }\href {https://doi.org/10.1103/PhysRevLett.132.236401} {\bibfield  {journal} {\bibinfo  {journal} {Phys. Rev. Lett.}\ }\textbf {\bibinfo {volume} {132}},\ \bibinfo {pages} {236401} (\bibinfo {year} {2024})}\BibitemShut {NoStop}%
\bibitem [{\citenamefont {Xiao}\ and\ \citenamefont {Zhao}(2024)}]{XiaoRevealing2024}%
  \BibitemOpen
  \bibfield  {author} {\bibinfo {author} {\bibfnamefont {R.}~\bibnamefont {Xiao}}\ and\ \bibinfo {author} {\bibfnamefont {Y.~X.}\ \bibnamefont {Zhao}},\ }\bibfield  {title} {\bibinfo {title} {Revealing the spatial nature of sublattice symmetry},\ }\href {https://doi.org/10.1038/s41467-024-48170-y} {\bibfield  {journal} {\bibinfo  {journal} {Nat. Commun.}\ }\textbf {\bibinfo {volume} {15}},\ \bibinfo {pages} {3787} (\bibinfo {year} {2024})}\BibitemShut {NoStop}%
\bibitem [{\citenamefont {Hu}\ \emph {et~al.}(2024)\citenamefont {Hu}, \citenamefont {Zhuang},\ and\ \citenamefont {Yang}}]{HuHigherOrder2024}%
  \BibitemOpen
  \bibfield  {author} {\bibinfo {author} {\bibfnamefont {J.}~\bibnamefont {Hu}}, \bibinfo {author} {\bibfnamefont {S.}~\bibnamefont {Zhuang}},\ and\ \bibinfo {author} {\bibfnamefont {Y.}~\bibnamefont {Yang}},\ }\bibfield  {title} {\bibinfo {title} {Higher-{{Order Topological Insulators}} via {{Momentum-Space Nonsymmorphic Symmetries}}},\ }\href {https://doi.org/10.1103/PhysRevLett.132.213801} {\bibfield  {journal} {\bibinfo  {journal} {Phys. Rev. Lett.}\ }\textbf {\bibinfo {volume} {132}},\ \bibinfo {pages} {213801} (\bibinfo {year} {2024})}\BibitemShut {NoStop}%
\bibitem [{\citenamefont {Fonseca}\ \emph {et~al.}(2024)\citenamefont {Fonseca}, \citenamefont {Vaidya}, \citenamefont {Christensen}, \citenamefont {Rechtsman}, \citenamefont {Hughes},\ and\ \citenamefont {Solja{\v c}i{\'c}}}]{FonsecaWeyl2024}%
  \BibitemOpen
  \bibfield  {author} {\bibinfo {author} {\bibfnamefont {A.~G.}\ \bibnamefont {Fonseca}}, \bibinfo {author} {\bibfnamefont {S.}~\bibnamefont {Vaidya}}, \bibinfo {author} {\bibfnamefont {T.}~\bibnamefont {Christensen}}, \bibinfo {author} {\bibfnamefont {M.~C.}\ \bibnamefont {Rechtsman}}, \bibinfo {author} {\bibfnamefont {T.~L.}\ \bibnamefont {Hughes}},\ and\ \bibinfo {author} {\bibfnamefont {M.}~\bibnamefont {Solja{\v c}i{\'c}}},\ }\bibfield  {title} {\bibinfo {title} {Weyl {{Points}} on {{Nonorientable Manifolds}}},\ }\href {https://doi.org/10.1103/PhysRevLett.132.266601} {\bibfield  {journal} {\bibinfo  {journal} {Phys. Rev. Lett.}\ }\textbf {\bibinfo {volume} {132}},\ \bibinfo {pages} {266601} (\bibinfo {year} {2024})}\BibitemShut {NoStop}%
\bibitem [{\citenamefont {Pu}\ \emph {et~al.}(2023)\citenamefont {Pu}, \citenamefont {He}, \citenamefont {Deng}, \citenamefont {Huang}, \citenamefont {Ye}, \citenamefont {Lu}, \citenamefont {Ke},\ and\ \citenamefont {Liu}}]{PuAcoustic2023}%
  \BibitemOpen
  \bibfield  {author} {\bibinfo {author} {\bibfnamefont {Z.}~\bibnamefont {Pu}}, \bibinfo {author} {\bibfnamefont {H.}~\bibnamefont {He}}, \bibinfo {author} {\bibfnamefont {W.}~\bibnamefont {Deng}}, \bibinfo {author} {\bibfnamefont {X.}~\bibnamefont {Huang}}, \bibinfo {author} {\bibfnamefont {L.}~\bibnamefont {Ye}}, \bibinfo {author} {\bibfnamefont {J.}~\bibnamefont {Lu}}, \bibinfo {author} {\bibfnamefont {M.}~\bibnamefont {Ke}},\ and\ \bibinfo {author} {\bibfnamefont {Z.}~\bibnamefont {Liu}},\ }\bibfield  {title} {\bibinfo {title} {Acoustic {{Klein}} bottle insulator},\ }\href {https://doi.org/10.1103/PhysRevB.108.L220101} {\bibfield  {journal} {\bibinfo  {journal} {Phys. Rev. B}\ }\textbf {\bibinfo {volume} {108}},\ \bibinfo {pages} {L220101} (\bibinfo {year} {2023})}\BibitemShut {NoStop}%
\bibitem [{\citenamefont {Wang}\ \emph {et~al.}(2023)\citenamefont {Wang}, \citenamefont {Zhang}, \citenamefont {Chen}, \citenamefont {Liang}, \citenamefont {Zhao},\ and\ \citenamefont {Cheng}}]{WangChessboard2023}%
  \BibitemOpen
  \bibfield  {author} {\bibinfo {author} {\bibfnamefont {Y.}~\bibnamefont {Wang}}, \bibinfo {author} {\bibfnamefont {C.}~\bibnamefont {Zhang}}, \bibinfo {author} {\bibfnamefont {Z.~Y.}\ \bibnamefont {Chen}}, \bibinfo {author} {\bibfnamefont {B.}~\bibnamefont {Liang}}, \bibinfo {author} {\bibfnamefont {Y.~X.}\ \bibnamefont {Zhao}},\ and\ \bibinfo {author} {\bibfnamefont {J.}~\bibnamefont {Cheng}},\ }\href {https://doi.org/10.48550/arXiv.2305.07174} {\bibinfo {title} {Chess-board {{Acoustic Crystals}} with {{Momentum-space Nonsymmorphic Symmetries}}}} (\bibinfo {year} {2023}),\ \Eprint {https://arxiv.org/abs/2305.07174} {arXiv:2305.07174 [cond-mat.mes-hall]} \BibitemShut {NoStop}%
\bibitem [{\citenamefont {Tao}\ \emph {et~al.}(2024)\citenamefont {Tao}, \citenamefont {Yan}, \citenamefont {Peng}, \citenamefont {Wei}, \citenamefont {Cui}, \citenamefont {Yang}, \citenamefont {Chen},\ and\ \citenamefont {Xu}}]{TaoHigherorder2024}%
  \BibitemOpen
  \bibfield  {author} {\bibinfo {author} {\bibfnamefont {Y.-L.}\ \bibnamefont {Tao}}, \bibinfo {author} {\bibfnamefont {M.}~\bibnamefont {Yan}}, \bibinfo {author} {\bibfnamefont {M.}~\bibnamefont {Peng}}, \bibinfo {author} {\bibfnamefont {Q.}~\bibnamefont {Wei}}, \bibinfo {author} {\bibfnamefont {Z.}~\bibnamefont {Cui}}, \bibinfo {author} {\bibfnamefont {S.~A.}\ \bibnamefont {Yang}}, \bibinfo {author} {\bibfnamefont {G.}~\bibnamefont {Chen}},\ and\ \bibinfo {author} {\bibfnamefont {Y.}~\bibnamefont {Xu}},\ }\bibfield  {title} {\bibinfo {title} {Higher-order {{Klein}} bottle topological insulator in three-dimensional acoustic crystals},\ }\href {https://doi.org/10.1103/PhysRevB.109.134107} {\bibfield  {journal} {\bibinfo  {journal} {Phys. Rev. B}\ }\textbf {\bibinfo {volume} {109}},\ \bibinfo {pages} {134107} (\bibinfo {year} {2024})}\BibitemShut {NoStop}%
\bibitem [{\citenamefont {Zhu}\ \emph {et~al.}(2024)\citenamefont {Zhu}, \citenamefont {Yang}, \citenamefont {Wu}, \citenamefont {Meng}, \citenamefont {Xi}, \citenamefont {Yan}, \citenamefont {Chen}, \citenamefont {Lu}, \citenamefont {Huang}, \citenamefont {Deng}, \citenamefont {Shang}, \citenamefont {Shum}, \citenamefont {Yang}, \citenamefont {Chen}, \citenamefont {Xiang}, \citenamefont {Liu}, \citenamefont {Liu},\ and\ \citenamefont {Gao}}]{ZhuBrillouin2024}%
  \BibitemOpen
  \bibfield  {author} {\bibinfo {author} {\bibfnamefont {Z.}~\bibnamefont {Zhu}}, \bibinfo {author} {\bibfnamefont {L.}~\bibnamefont {Yang}}, \bibinfo {author} {\bibfnamefont {J.}~\bibnamefont {Wu}}, \bibinfo {author} {\bibfnamefont {Y.}~\bibnamefont {Meng}}, \bibinfo {author} {\bibfnamefont {X.}~\bibnamefont {Xi}}, \bibinfo {author} {\bibfnamefont {B.}~\bibnamefont {Yan}}, \bibinfo {author} {\bibfnamefont {J.}~\bibnamefont {Chen}}, \bibinfo {author} {\bibfnamefont {J.}~\bibnamefont {Lu}}, \bibinfo {author} {\bibfnamefont {X.}~\bibnamefont {Huang}}, \bibinfo {author} {\bibfnamefont {W.}~\bibnamefont {Deng}}, \bibinfo {author} {\bibfnamefont {C.}~\bibnamefont {Shang}}, \bibinfo {author} {\bibfnamefont {P.~P.}\ \bibnamefont {Shum}}, \bibinfo {author} {\bibfnamefont {Y.}~\bibnamefont {Yang}}, \bibinfo {author} {\bibfnamefont {H.}~\bibnamefont {Chen}}, \bibinfo {author} {\bibfnamefont {K.}~\bibnamefont {Xiang}}, \bibinfo {author} {\bibfnamefont {G.-G.}\ \bibnamefont {Liu}}, \bibinfo {author} {\bibfnamefont {Z.}~\bibnamefont {Liu}},\ and\ \bibinfo {author} {\bibfnamefont {Z.}~\bibnamefont {Gao}},\ }\bibfield  {title} {\bibinfo {title} {Brillouin {{Klein}} space and half-turn space in three-dimensional acoustic crystals},\ }\href {https://doi.org/10.1016/j.scib.2024.05.003} {\bibfield  {journal} {\bibinfo  {journal} {Sci. Bull.}\ }\textbf {\bibinfo {volume} {69}},\ \bibinfo {pages} {2050} (\bibinfo {year} {2024})}\BibitemShut {NoStop}%
\bibitem [{\citenamefont {Lai}\ \emph {et~al.}(2024)\citenamefont {Lai}, \citenamefont {Wu}, \citenamefont {Pu}, \citenamefont {Zhou}, \citenamefont {Lu}, \citenamefont {Liu}, \citenamefont {Deng}, \citenamefont {Cheng}, \citenamefont {Chen},\ and\ \citenamefont {Liu}}]{LaiRealprojectiveplane2024}%
  \BibitemOpen
  \bibfield  {author} {\bibinfo {author} {\bibfnamefont {P.}~\bibnamefont {Lai}}, \bibinfo {author} {\bibfnamefont {J.}~\bibnamefont {Wu}}, \bibinfo {author} {\bibfnamefont {Z.}~\bibnamefont {Pu}}, \bibinfo {author} {\bibfnamefont {Q.}~\bibnamefont {Zhou}}, \bibinfo {author} {\bibfnamefont {J.}~\bibnamefont {Lu}}, \bibinfo {author} {\bibfnamefont {H.}~\bibnamefont {Liu}}, \bibinfo {author} {\bibfnamefont {W.}~\bibnamefont {Deng}}, \bibinfo {author} {\bibfnamefont {H.}~\bibnamefont {Cheng}}, \bibinfo {author} {\bibfnamefont {S.}~\bibnamefont {Chen}},\ and\ \bibinfo {author} {\bibfnamefont {Z.}~\bibnamefont {Liu}},\ }\bibfield  {title} {\bibinfo {title} {Real-projective-plane hybrid-order topological insulator realized in phononic crystals},\ }\href {https://doi.org/10.1103/PhysRevApplied.21.044002} {\bibfield  {journal} {\bibinfo  {journal} {Phys. Rev. Appl.}\ }\textbf {\bibinfo {volume} {21}},\ \bibinfo {pages} {044002} (\bibinfo {year} {2024})}\BibitemShut {NoStop}%
\bibitem [{\citenamefont {Liu}\ \emph {et~al.}(2024)\citenamefont {Liu}, \citenamefont {Jiang}, \citenamefont {Wen}, \citenamefont {Song}, \citenamefont {Li}, \citenamefont {Lu},\ and\ \citenamefont {Ke}}]{LiuTopological2024}%
  \BibitemOpen
  \bibfield  {author} {\bibinfo {author} {\bibfnamefont {Y.}~\bibnamefont {Liu}}, \bibinfo {author} {\bibfnamefont {C.}~\bibnamefont {Jiang}}, \bibinfo {author} {\bibfnamefont {W.}~\bibnamefont {Wen}}, \bibinfo {author} {\bibfnamefont {Y.}~\bibnamefont {Song}}, \bibinfo {author} {\bibfnamefont {X.}~\bibnamefont {Li}}, \bibinfo {author} {\bibfnamefont {P.}~\bibnamefont {Lu}},\ and\ \bibinfo {author} {\bibfnamefont {S.}~\bibnamefont {Ke}},\ }\bibfield  {title} {\bibinfo {title} {Topological phases in photonic microring lattices with projective symmetry},\ }\href {https://doi.org/10.1103/PhysRevA.109.013516} {\bibfield  {journal} {\bibinfo  {journal} {Phys. Rev. A}\ }\textbf {\bibinfo {volume} {109}},\ \bibinfo {pages} {013516} (\bibinfo {year} {2024})}\BibitemShut {NoStop}%
\bibitem [{\citenamefont {Zhang}\ \emph {et~al.}(2025)\citenamefont {Zhang}, \citenamefont {Wang}, \citenamefont {Lyu},\ and\ \citenamefont {Zhao}}]{ZhangBrillouin2025}%
  \BibitemOpen
  \bibfield  {author} {\bibinfo {author} {\bibfnamefont {C.}~\bibnamefont {Zhang}}, \bibinfo {author} {\bibfnamefont {P.}~\bibnamefont {Wang}}, \bibinfo {author} {\bibfnamefont {J.}~\bibnamefont {Lyu}},\ and\ \bibinfo {author} {\bibfnamefont {Y.~X.}\ \bibnamefont {Zhao}},\ }\bibfield  {title} {\bibinfo {title} {Brillouin {{Platycosms}} and {{Topological Phases}}},\ }\href {https://doi.org/10.1103/pllk-n4h3} {\bibfield  {journal} {\bibinfo  {journal} {Phys. Rev. Lett.}\ }\textbf {\bibinfo {volume} {135}},\ \bibinfo {pages} {136601} (\bibinfo {year} {2025})}\BibitemShut {NoStop}%
\bibitem [{\citenamefont {D{\"u}rrnagel}\ \emph {et~al.}(2026{\natexlab{a}})\citenamefont {D{\"u}rrnagel}, \citenamefont {Klebl}, \citenamefont {M{\"u}ller}, \citenamefont {Thomale},\ and\ \citenamefont {Klett}}]{DurrnagelExtended2026}%
  \BibitemOpen
  \bibfield  {author} {\bibinfo {author} {\bibfnamefont {M.}~\bibnamefont {D{\"u}rrnagel}}, \bibinfo {author} {\bibfnamefont {L.}~\bibnamefont {Klebl}}, \bibinfo {author} {\bibfnamefont {T.}~\bibnamefont {M{\"u}ller}}, \bibinfo {author} {\bibfnamefont {R.}~\bibnamefont {Thomale}},\ and\ \bibinfo {author} {\bibfnamefont {M.}~\bibnamefont {Klett}},\ }\href {https://doi.org/10.48550/arXiv.2508.20163} {\bibinfo {title} {Extended s-wave altermagnets}} (\bibinfo {year} {2026}{\natexlab{a}}),\ \Eprint {https://arxiv.org/abs/2508.20163} {arXiv:2508.20163 [cond-mat.str-el]} \BibitemShut {NoStop}%
\bibitem [{\citenamefont {D{\"u}rrnagel}\ \emph {et~al.}(2026{\natexlab{b}})\citenamefont {D{\"u}rrnagel}, \citenamefont {Baum}, \citenamefont {Klett}, \citenamefont {Klebl},\ and\ \citenamefont {Thomale}}]{DurrnagelPair2026}%
  \BibitemOpen
  \bibfield  {author} {\bibinfo {author} {\bibfnamefont {M.}~\bibnamefont {D{\"u}rrnagel}}, \bibinfo {author} {\bibfnamefont {C.~A.}\ \bibnamefont {Baum}}, \bibinfo {author} {\bibfnamefont {M.}~\bibnamefont {Klett}}, \bibinfo {author} {\bibfnamefont {L.}~\bibnamefont {Klebl}},\ and\ \bibinfo {author} {\bibfnamefont {R.}~\bibnamefont {Thomale}},\ }\href {https://doi.org/10.48550/arXiv.2609.05603} {\bibinfo {title} {Pair density wave order from non-symmorphic momentum symmetry}} (\bibinfo {year} {2026}{\natexlab{b}}),\ \Eprint {https://arxiv.org/abs/2609.05603} {arXiv:2609.05603 [cond-mat.supr-con]} \BibitemShut {NoStop}%
\bibitem [{\citenamefont {Xiao}\ \emph {et~al.}(2024)\citenamefont {Xiao}, \citenamefont {Zhao}, \citenamefont {Li}, \citenamefont {Shindou},\ and\ \citenamefont {Song}}]{XiaoSpin2024}%
  \BibitemOpen
  \bibfield  {author} {\bibinfo {author} {\bibfnamefont {Z.}~\bibnamefont {Xiao}}, \bibinfo {author} {\bibfnamefont {J.}~\bibnamefont {Zhao}}, \bibinfo {author} {\bibfnamefont {Y.}~\bibnamefont {Li}}, \bibinfo {author} {\bibfnamefont {R.}~\bibnamefont {Shindou}},\ and\ \bibinfo {author} {\bibfnamefont {Z.-D.}\ \bibnamefont {Song}},\ }\bibfield  {title} {\bibinfo {title} {Spin {{Space Groups}}: {{Full Classification}} and {{Applications}}},\ }\href {https://doi.org/10.1103/PhysRevX.14.031037} {\bibfield  {journal} {\bibinfo  {journal} {Phys. Rev. X}\ }\textbf {\bibinfo {volume} {14}},\ \bibinfo {pages} {031037} (\bibinfo {year} {2024})}\BibitemShut {NoStop}%
\bibitem [{\citenamefont {Brinkman}\ and\ \citenamefont {Elliott}(1966)}]{BrinkmanTheory1966}%
  \BibitemOpen
  \bibfield  {author} {\bibinfo {author} {\bibfnamefont {W.~F.}\ \bibnamefont {Brinkman}}\ and\ \bibinfo {author} {\bibfnamefont {R.~J.}\ \bibnamefont {Elliott}},\ }\bibfield  {title} {\bibinfo {title} {Theory of {{Spin-Space Groups}}},\ }\href@noop {} {\bibfield  {journal} {\bibinfo  {journal} {Proc. R. Soc. A}\ }\textbf {\bibinfo {volume} {294}},\ \bibinfo {pages} {343} (\bibinfo {year} {1966})}\BibitemShut {NoStop}%
\bibitem [{\citenamefont {Litvin}\ and\ \citenamefont {Opechowski}(1974)}]{LitvinSpin1974}%
  \BibitemOpen
  \bibfield  {author} {\bibinfo {author} {\bibfnamefont {D.~B.}\ \bibnamefont {Litvin}}\ and\ \bibinfo {author} {\bibfnamefont {W.}~\bibnamefont {Opechowski}},\ }\bibfield  {title} {\bibinfo {title} {Spin groups},\ }\href {https://doi.org/10.1016/0031-8914(74)90157-8} {\bibfield  {journal} {\bibinfo  {journal} {Physica}\ }\textbf {\bibinfo {volume} {76}},\ \bibinfo {pages} {538} (\bibinfo {year} {1974})}\BibitemShut {NoStop}%
\bibitem [{\citenamefont {Sandratskii}(1998)}]{SandratskiiNoncollinear1998}%
  \BibitemOpen
  \bibfield  {author} {\bibinfo {author} {\bibfnamefont {L.~M.}\ \bibnamefont {Sandratskii}},\ }\bibfield  {title} {\bibinfo {title} {Noncollinear magnetism in itinerant-electron systems: {{Theory}} and applications},\ }\href {https://doi.org/10.1080/000187398243573} {\bibfield  {journal} {\bibinfo  {journal} {Adv. Phys.}\ }\textbf {\bibinfo {volume} {47}},\ \bibinfo {pages} {91} (\bibinfo {year} {1998})}\BibitemShut {NoStop}%
\bibitem [{\citenamefont {Liu}\ \emph {et~al.}(2022)\citenamefont {Liu}, \citenamefont {Li}, \citenamefont {Han}, \citenamefont {Wan},\ and\ \citenamefont {Liu}}]{LiuSpinGroup2022}%
  \BibitemOpen
  \bibfield  {author} {\bibinfo {author} {\bibfnamefont {P.}~\bibnamefont {Liu}}, \bibinfo {author} {\bibfnamefont {J.}~\bibnamefont {Li}}, \bibinfo {author} {\bibfnamefont {J.}~\bibnamefont {Han}}, \bibinfo {author} {\bibfnamefont {X.}~\bibnamefont {Wan}},\ and\ \bibinfo {author} {\bibfnamefont {Q.}~\bibnamefont {Liu}},\ }\bibfield  {title} {\bibinfo {title} {Spin-{{Group Symmetry}} in {{Magnetic Materials}} with {{Negligible Spin-Orbit Coupling}}},\ }\href {https://doi.org/10.1103/PhysRevX.12.021016} {\bibfield  {journal} {\bibinfo  {journal} {Phys. Rev. X}\ }\textbf {\bibinfo {volume} {12}},\ \bibinfo {pages} {021016} (\bibinfo {year} {2022})}\BibitemShut {NoStop}%
\bibitem [{\citenamefont {{\v S}mejkal}\ \emph {et~al.}(2022{\natexlab{a}})\citenamefont {{\v S}mejkal}, \citenamefont {Sinova},\ and\ \citenamefont {Jungwirth}}]{SmejkalConventional2022}%
  \BibitemOpen
  \bibfield  {author} {\bibinfo {author} {\bibfnamefont {L.}~\bibnamefont {{\v S}mejkal}}, \bibinfo {author} {\bibfnamefont {J.}~\bibnamefont {Sinova}},\ and\ \bibinfo {author} {\bibfnamefont {T.}~\bibnamefont {Jungwirth}},\ }\bibfield  {title} {\bibinfo {title} {Beyond {{Conventional Ferromagnetism}} and {{Antiferromagnetism}}: {{A Phase}} with {{Nonrelativistic Spin}} and {{Crystal Rotation Symmetry}}},\ }\href {https://doi.org/10.1103/PhysRevX.12.031042} {\bibfield  {journal} {\bibinfo  {journal} {Phys. Rev. X}\ }\textbf {\bibinfo {volume} {12}},\ \bibinfo {pages} {031042} (\bibinfo {year} {2022}{\natexlab{a}})}\BibitemShut {NoStop}%
\bibitem [{\citenamefont {{\v S}mejkal}\ \emph {et~al.}(2022{\natexlab{b}})\citenamefont {{\v S}mejkal}, \citenamefont {Sinova},\ and\ \citenamefont {Jungwirth}}]{SmejkalEmerging2022}%
  \BibitemOpen
  \bibfield  {author} {\bibinfo {author} {\bibfnamefont {L.}~\bibnamefont {{\v S}mejkal}}, \bibinfo {author} {\bibfnamefont {J.}~\bibnamefont {Sinova}},\ and\ \bibinfo {author} {\bibfnamefont {T.}~\bibnamefont {Jungwirth}},\ }\bibfield  {title} {\bibinfo {title} {Emerging {{Research Landscape}} of {{Altermagnetism}}},\ }\href {https://doi.org/10.1103/PhysRevX.12.040501} {\bibfield  {journal} {\bibinfo  {journal} {Phys. Rev. X}\ }\textbf {\bibinfo {volume} {12}},\ \bibinfo {pages} {040501} (\bibinfo {year} {2022}{\natexlab{b}})}\BibitemShut {NoStop}%
\bibitem [{\citenamefont {Corticelli}\ \emph {et~al.}(2022)\citenamefont {Corticelli}, \citenamefont {Moessner},\ and\ \citenamefont {McClarty}}]{CorticelliSpinspace2022}%
  \BibitemOpen
  \bibfield  {author} {\bibinfo {author} {\bibfnamefont {A.}~\bibnamefont {Corticelli}}, \bibinfo {author} {\bibfnamefont {R.}~\bibnamefont {Moessner}},\ and\ \bibinfo {author} {\bibfnamefont {P.~A.}\ \bibnamefont {McClarty}},\ }\bibfield  {title} {\bibinfo {title} {Spin-space groups and magnon band topology},\ }\href {https://doi.org/10.1103/PhysRevB.105.064430} {\bibfield  {journal} {\bibinfo  {journal} {Phys. Rev. B}\ }\textbf {\bibinfo {volume} {105}},\ \bibinfo {pages} {064430} (\bibinfo {year} {2022})}\BibitemShut {NoStop}%
\bibitem [{\citenamefont {Zeng}\ and\ \citenamefont {Zhao}(2024)}]{ZengDescription2024}%
  \BibitemOpen
  \bibfield  {author} {\bibinfo {author} {\bibfnamefont {S.}~\bibnamefont {Zeng}}\ and\ \bibinfo {author} {\bibfnamefont {Y.-J.}\ \bibnamefont {Zhao}},\ }\bibfield  {title} {\bibinfo {title} {Description of two-dimensional altermagnetism: {{Categorization}} using spin group theory},\ }\href {https://doi.org/10.1103/PhysRevB.110.054406} {\bibfield  {journal} {\bibinfo  {journal} {Phys. Rev. B}\ }\textbf {\bibinfo {volume} {110}},\ \bibinfo {pages} {054406} (\bibinfo {year} {2024})}\BibitemShut {NoStop}%
\bibitem [{\citenamefont {Chen}\ \emph {et~al.}(2024)\citenamefont {Chen}, \citenamefont {Ren}, \citenamefont {Zhu}, \citenamefont {Yu}, \citenamefont {Zhang}, \citenamefont {Liu}, \citenamefont {Li}, \citenamefont {Liu}, \citenamefont {Li},\ and\ \citenamefont {Liu}}]{ChenEnumeration2024}%
  \BibitemOpen
  \bibfield  {author} {\bibinfo {author} {\bibfnamefont {X.}~\bibnamefont {Chen}}, \bibinfo {author} {\bibfnamefont {J.}~\bibnamefont {Ren}}, \bibinfo {author} {\bibfnamefont {Y.}~\bibnamefont {Zhu}}, \bibinfo {author} {\bibfnamefont {Y.}~\bibnamefont {Yu}}, \bibinfo {author} {\bibfnamefont {A.}~\bibnamefont {Zhang}}, \bibinfo {author} {\bibfnamefont {P.}~\bibnamefont {Liu}}, \bibinfo {author} {\bibfnamefont {J.}~\bibnamefont {Li}}, \bibinfo {author} {\bibfnamefont {Y.}~\bibnamefont {Liu}}, \bibinfo {author} {\bibfnamefont {C.}~\bibnamefont {Li}},\ and\ \bibinfo {author} {\bibfnamefont {Q.}~\bibnamefont {Liu}},\ }\bibfield  {title} {\bibinfo {title} {Enumeration and {{Representation Theory}} of {{Spin Space Groups}}},\ }\href {https://doi.org/10.1103/PhysRevX.14.031038} {\bibfield  {journal} {\bibinfo  {journal} {Phys. Rev. X}\ }\textbf {\bibinfo {volume} {14}},\ \bibinfo {pages} {031038} (\bibinfo {year} {2024})}\BibitemShut {NoStop}%
\bibitem [{\citenamefont {Jiang}\ \emph {et~al.}(2024)\citenamefont {Jiang}, \citenamefont {Song}, \citenamefont {Zhu}, \citenamefont {Fang}, \citenamefont {Weng}, \citenamefont {Liu}, \citenamefont {Yang},\ and\ \citenamefont {Fang}}]{JiangEnumeration2024}%
  \BibitemOpen
  \bibfield  {author} {\bibinfo {author} {\bibfnamefont {Y.}~\bibnamefont {Jiang}}, \bibinfo {author} {\bibfnamefont {Z.}~\bibnamefont {Song}}, \bibinfo {author} {\bibfnamefont {T.}~\bibnamefont {Zhu}}, \bibinfo {author} {\bibfnamefont {Z.}~\bibnamefont {Fang}}, \bibinfo {author} {\bibfnamefont {H.}~\bibnamefont {Weng}}, \bibinfo {author} {\bibfnamefont {Z.-X.}\ \bibnamefont {Liu}}, \bibinfo {author} {\bibfnamefont {J.}~\bibnamefont {Yang}},\ and\ \bibinfo {author} {\bibfnamefont {C.}~\bibnamefont {Fang}},\ }\bibfield  {title} {\bibinfo {title} {Enumeration of {{Spin-Space Groups}}: {{Toward}} a {{Complete Description}} of {{Symmetries}} of {{Magnetic Orders}}},\ }\href {https://doi.org/10.1103/PhysRevX.14.031039} {\bibfield  {journal} {\bibinfo  {journal} {Phys. Rev. X}\ }\textbf {\bibinfo {volume} {14}},\ \bibinfo {pages} {031039} (\bibinfo {year} {2024})}\BibitemShut {NoStop}%
\bibitem [{\citenamefont {Radaelli}(2024)}]{RadaelliTensorial2024}%
  \BibitemOpen
  \bibfield  {author} {\bibinfo {author} {\bibfnamefont {P.~G.}\ \bibnamefont {Radaelli}},\ }\bibfield  {title} {\bibinfo {title} {Tensorial approach to altermagnetism},\ }\href {https://doi.org/10.1103/PhysRevB.110.214428} {\bibfield  {journal} {\bibinfo  {journal} {Phys. Rev. B}\ }\textbf {\bibinfo {volume} {110}},\ \bibinfo {pages} {214428} (\bibinfo {year} {2024})}\BibitemShut {NoStop}%
\bibitem [{\citenamefont {Chen}\ \emph {et~al.}(2025)\citenamefont {Chen}, \citenamefont {Liu}, \citenamefont {Liu}, \citenamefont {Yu}, \citenamefont {Ren}, \citenamefont {Li}, \citenamefont {Zhang},\ and\ \citenamefont {Liu}}]{ChenUnconventional2025}%
  \BibitemOpen
  \bibfield  {author} {\bibinfo {author} {\bibfnamefont {X.}~\bibnamefont {Chen}}, \bibinfo {author} {\bibfnamefont {Y.}~\bibnamefont {Liu}}, \bibinfo {author} {\bibfnamefont {P.}~\bibnamefont {Liu}}, \bibinfo {author} {\bibfnamefont {Y.}~\bibnamefont {Yu}}, \bibinfo {author} {\bibfnamefont {J.}~\bibnamefont {Ren}}, \bibinfo {author} {\bibfnamefont {J.}~\bibnamefont {Li}}, \bibinfo {author} {\bibfnamefont {A.}~\bibnamefont {Zhang}},\ and\ \bibinfo {author} {\bibfnamefont {Q.}~\bibnamefont {Liu}},\ }\bibfield  {title} {\bibinfo {title} {Unconventional magnons in collinear magnets dictated by spin space groups},\ }\href {https://doi.org/10.1038/s41586-025-08715-7} {\bibfield  {journal} {\bibinfo  {journal} {Nature}\ }\textbf {\bibinfo {volume} {640}},\ \bibinfo {pages} {349} (\bibinfo {year} {2025})}\BibitemShut {NoStop}%
\bibitem [{\citenamefont {Biniskos}\ \emph {et~al.}(2025)\citenamefont {Biniskos}, \citenamefont {{dos Santos Dias}}, \citenamefont {Agrestini}, \citenamefont {Svit{\'a}k}, \citenamefont {Zhou}, \citenamefont {Posp{\'i}{\v s}il},\ and\ \citenamefont {{\v C}erm{\'a}k}}]{BiniskosSystematic2025}%
  \BibitemOpen
  \bibfield  {author} {\bibinfo {author} {\bibfnamefont {N.}~\bibnamefont {Biniskos}}, \bibinfo {author} {\bibfnamefont {M.}~\bibnamefont {{dos Santos Dias}}}, \bibinfo {author} {\bibfnamefont {S.}~\bibnamefont {Agrestini}}, \bibinfo {author} {\bibfnamefont {D.}~\bibnamefont {Svit{\'a}k}}, \bibinfo {author} {\bibfnamefont {K.-J.}\ \bibnamefont {Zhou}}, \bibinfo {author} {\bibfnamefont {J.}~\bibnamefont {Posp{\'i}{\v s}il}},\ and\ \bibinfo {author} {\bibfnamefont {P.}~\bibnamefont {{\v C}erm{\'a}k}},\ }\bibfield  {title} {\bibinfo {title} {Systematic mapping of altermagnetic magnons by resonant inelastic {{X-ray}} circular dichroism},\ }\href {https://doi.org/10.1038/s41467-025-64322-0} {\bibfield  {journal} {\bibinfo  {journal} {Nat. Commun.}\ }\textbf {\bibinfo {volume} {16}},\ \bibinfo {pages} {9311} (\bibinfo {year} {2025})}\BibitemShut {NoStop}%
\bibitem [{\citenamefont {Feng}\ and\ \citenamefont {Zhang}(2025)}]{FengSuperconducting2025}%
  \BibitemOpen
  \bibfield  {author} {\bibinfo {author} {\bibfnamefont {X.}~\bibnamefont {Feng}}\ and\ \bibinfo {author} {\bibfnamefont {Z.}~\bibnamefont {Zhang}},\ }\bibfield  {title} {\bibinfo {title} {Superconducting order parameters in spin space groups: {{Methodology}} and application},\ }\href {https://doi.org/10.1103/PhysRevB.111.054520} {\bibfield  {journal} {\bibinfo  {journal} {Phys. Rev. B}\ }\textbf {\bibinfo {volume} {111}},\ \bibinfo {pages} {054520} (\bibinfo {year} {2025})}\BibitemShut {NoStop}%
\bibitem [{\citenamefont {Devaraj}\ \emph {et~al.}(2024)\citenamefont {Devaraj}, \citenamefont {Bose},\ and\ \citenamefont {Narayan}}]{DevarajInterplay2024}%
  \BibitemOpen
  \bibfield  {author} {\bibinfo {author} {\bibfnamefont {N.}~\bibnamefont {Devaraj}}, \bibinfo {author} {\bibfnamefont {A.}~\bibnamefont {Bose}},\ and\ \bibinfo {author} {\bibfnamefont {A.}~\bibnamefont {Narayan}},\ }\bibfield  {title} {\bibinfo {title} {Interplay of altermagnetism and pressure in hexagonal and orthorhombic {{MnTe}}},\ }\href {https://doi.org/10.1103/PhysRevMaterials.8.104407} {\bibfield  {journal} {\bibinfo  {journal} {Phys. Rev. Mater.}\ }\textbf {\bibinfo {volume} {8}},\ \bibinfo {pages} {104407} (\bibinfo {year} {2024})}\BibitemShut {NoStop}%
\bibitem [{\citenamefont {Etxebarria}\ \emph {et~al.}(2025)\citenamefont {Etxebarria}, \citenamefont {{Perez-Mato}}, \citenamefont {Tasci},\ and\ \citenamefont {Elcoro}}]{EtxebarriaCrystal2025}%
  \BibitemOpen
  \bibfield  {author} {\bibinfo {author} {\bibfnamefont {J.}~\bibnamefont {Etxebarria}}, \bibinfo {author} {\bibfnamefont {J.~M.}\ \bibnamefont {{Perez-Mato}}}, \bibinfo {author} {\bibfnamefont {E.~S.}\ \bibnamefont {Tasci}},\ and\ \bibinfo {author} {\bibfnamefont {L.}~\bibnamefont {Elcoro}},\ }\bibfield  {title} {\bibinfo {title} {Crystal tensor properties of magnetic materials with and without spin--orbit coupling. {{Application}} of spin point groups as approximate symmetries},\ }\href {https://doi.org/10.1107/S2053273325004127} {\bibfield  {journal} {\bibinfo  {journal} {Acta Cryst A}\ }\textbf {\bibinfo {volume} {81}},\ \bibinfo {pages} {317} (\bibinfo {year} {2025})}\BibitemShut {NoStop}%
\bibitem [{\citenamefont {Elcoro}\ \emph {et~al.}(2026)\citenamefont {Elcoro}, \citenamefont {Etxebarria}, \citenamefont {{Perez-Mato}},\ and\ \citenamefont {Tasci}}]{ElcoroAutomatic2026}%
  \BibitemOpen
  \bibfield  {author} {\bibinfo {author} {\bibfnamefont {L.}~\bibnamefont {Elcoro}}, \bibinfo {author} {\bibfnamefont {J.}~\bibnamefont {Etxebarria}}, \bibinfo {author} {\bibfnamefont {J.~M.}\ \bibnamefont {{Perez-Mato}}},\ and\ \bibinfo {author} {\bibfnamefont {E.~S.}\ \bibnamefont {Tasci}},\ }\bibfield  {title} {\bibinfo {title} {Automatic calculation of symmetry-adapted tensors under spin-group symmetry: {{STENSOR}}, a new tool of the {{Bilbao Crystallographic Server}}},\ }\href {https://doi.org/10.1107/S1600576726000944} {\bibfield  {journal} {\bibinfo  {journal} {J. Appl. Cryst.}\ }\textbf {\bibinfo {volume} {59}},\ \bibinfo {pages} {640} (\bibinfo {year} {2026})}\BibitemShut {NoStop}%
\bibitem [{\citenamefont {Guo}\ \emph {et~al.}(2021)\citenamefont {Guo}, \citenamefont {Wei}, \citenamefont {Liu}, \citenamefont {Liu},\ and\ \citenamefont {Lu}}]{GuoEightfold2021}%
  \BibitemOpen
  \bibfield  {author} {\bibinfo {author} {\bibfnamefont {P.-J.}\ \bibnamefont {Guo}}, \bibinfo {author} {\bibfnamefont {Y.-W.}\ \bibnamefont {Wei}}, \bibinfo {author} {\bibfnamefont {K.}~\bibnamefont {Liu}}, \bibinfo {author} {\bibfnamefont {Z.-X.}\ \bibnamefont {Liu}},\ and\ \bibinfo {author} {\bibfnamefont {Z.-Y.}\ \bibnamefont {Lu}},\ }\bibfield  {title} {\bibinfo {title} {Eightfold {{Degenerate Fermions}} in {{Two Dimensions}}},\ }\href {https://doi.org/10.1103/PhysRevLett.127.176401} {\bibfield  {journal} {\bibinfo  {journal} {Phys. Rev. Lett.}\ }\textbf {\bibinfo {volume} {127}},\ \bibinfo {pages} {176401} (\bibinfo {year} {2021})}\BibitemShut {NoStop}%
\bibitem [{\citenamefont {Yang}\ \emph {et~al.}(2024)\citenamefont {Yang}, \citenamefont {Liu},\ and\ \citenamefont {Fang}}]{YangSymmetry2024}%
  \BibitemOpen
  \bibfield  {author} {\bibinfo {author} {\bibfnamefont {J.}~\bibnamefont {Yang}}, \bibinfo {author} {\bibfnamefont {Z.-X.}\ \bibnamefont {Liu}},\ and\ \bibinfo {author} {\bibfnamefont {C.}~\bibnamefont {Fang}},\ }\bibfield  {title} {\bibinfo {title} {Symmetry invariants and classes of quasiparticles in magnetically ordered systems having weak spin-orbit coupling},\ }\href {https://doi.org/10.1038/s41467-024-53862-6} {\bibfield  {journal} {\bibinfo  {journal} {Nat. Commun.}\ }\textbf {\bibinfo {volume} {15}},\ \bibinfo {pages} {10203} (\bibinfo {year} {2024})}\BibitemShut {NoStop}%
\bibitem [{\citenamefont {Li}\ \emph {et~al.}(2025)\citenamefont {Li}, \citenamefont {Hu}, \citenamefont {Li}, \citenamefont {Wang}, \citenamefont {Chen}, \citenamefont {Thiagarajan}, \citenamefont {Leandersson}, \citenamefont {Polley}, \citenamefont {Kim}, \citenamefont {Liu}, \citenamefont {Fulga}, \citenamefont {Vergniory}, \citenamefont {Janson}, \citenamefont {Tjernberg},\ and\ \citenamefont {{van den Brink}}}]{LiTopological2025}%
  \BibitemOpen
  \bibfield  {author} {\bibinfo {author} {\bibfnamefont {C.}~\bibnamefont {Li}}, \bibinfo {author} {\bibfnamefont {M.}~\bibnamefont {Hu}}, \bibinfo {author} {\bibfnamefont {Z.}~\bibnamefont {Li}}, \bibinfo {author} {\bibfnamefont {Y.}~\bibnamefont {Wang}}, \bibinfo {author} {\bibfnamefont {W.}~\bibnamefont {Chen}}, \bibinfo {author} {\bibfnamefont {B.}~\bibnamefont {Thiagarajan}}, \bibinfo {author} {\bibfnamefont {M.}~\bibnamefont {Leandersson}}, \bibinfo {author} {\bibfnamefont {C.}~\bibnamefont {Polley}}, \bibinfo {author} {\bibfnamefont {T.}~\bibnamefont {Kim}}, \bibinfo {author} {\bibfnamefont {H.}~\bibnamefont {Liu}}, \bibinfo {author} {\bibfnamefont {C.}~\bibnamefont {Fulga}}, \bibinfo {author} {\bibfnamefont {M.~G.}\ \bibnamefont {Vergniory}}, \bibinfo {author} {\bibfnamefont {O.}~\bibnamefont {Janson}}, \bibinfo {author} {\bibfnamefont {O.}~\bibnamefont {Tjernberg}},\ and\ \bibinfo {author} {\bibfnamefont {J.}~\bibnamefont {{van den Brink}}},\ }\bibfield  {title} {\bibinfo {title} {Topological {{Weyl}} altermagnetism in {{CrSb}}},\ }\href {https://doi.org/10.1038/s42005-025-02232-9} {\bibfield  {journal} {\bibinfo  {journal} {Commun. Phys.}\ }\textbf {\bibinfo {volume} {8}},\ \bibinfo {pages} {311} (\bibinfo {year} {2025})}\BibitemShut {NoStop}%
\bibitem [{\citenamefont {Yu}\ \emph {et~al.}(2025)\citenamefont {Yu}, \citenamefont {Shahid}, \citenamefont {Liu}, \citenamefont {Shao}, \citenamefont {Chen},\ and\ \citenamefont {Sun}}]{YuNeel2025}%
  \BibitemOpen
  \bibfield  {author} {\bibinfo {author} {\bibfnamefont {T.}~\bibnamefont {Yu}}, \bibinfo {author} {\bibfnamefont {I.}~\bibnamefont {Shahid}}, \bibinfo {author} {\bibfnamefont {P.}~\bibnamefont {Liu}}, \bibinfo {author} {\bibfnamefont {D.-F.}\ \bibnamefont {Shao}}, \bibinfo {author} {\bibfnamefont {X.-Q.}\ \bibnamefont {Chen}},\ and\ \bibinfo {author} {\bibfnamefont {Y.}~\bibnamefont {Sun}},\ }\bibfield  {title} {\bibinfo {title} {N\'eel vector-dependent anomalous transport in altermagnetic metal {{CrSb}}},\ }\href {https://doi.org/10.1038/s41535-025-00766-3} {\bibfield  {journal} {\bibinfo  {journal} {npj Quantum Mater.}\ }\textbf {\bibinfo {volume} {10}},\ \bibinfo {pages} {47} (\bibinfo {year} {2025})}\BibitemShut {NoStop}%
\bibitem [{\citenamefont {Takahashi}\ \emph {et~al.}(2025)\citenamefont {Takahashi}, \citenamefont {Huang}, \citenamefont {Yu},\ and\ \citenamefont {Zang}}]{TakahashiSymmetry2025}%
  \BibitemOpen
  \bibfield  {author} {\bibinfo {author} {\bibfnamefont {K.}~\bibnamefont {Takahashi}}, \bibinfo {author} {\bibfnamefont {H.-F.}\ \bibnamefont {Huang}}, \bibinfo {author} {\bibfnamefont {J.-X.}\ \bibnamefont {Yu}},\ and\ \bibinfo {author} {\bibfnamefont {J.}~\bibnamefont {Zang}},\ }\bibfield  {title} {\bibinfo {title} {Symmetry and minimal {{Hamiltonian}} of nonsymmorphic collinear antiferromagnet {{MnTe}}},\ }\href {https://doi.org/10.1038/s41535-025-00784-1} {\bibfield  {journal} {\bibinfo  {journal} {npj Quantum Mater.}\ }\textbf {\bibinfo {volume} {10}},\ \bibinfo {pages} {70} (\bibinfo {year} {2025})}\BibitemShut {NoStop}%
\bibitem [{\citenamefont {Wei{\ss}enhofer}\ \emph {et~al.}(2026)\citenamefont {Wei{\ss}enhofer}, \citenamefont {Mrudul}, \citenamefont {Mankovsky},\ and\ \citenamefont {Oppeneer}}]{WeissenhoferMagnon2026}%
  \BibitemOpen
  \bibfield  {author} {\bibinfo {author} {\bibfnamefont {M.}~\bibnamefont {Wei{\ss}enhofer}}, \bibinfo {author} {\bibfnamefont {M.~S.}\ \bibnamefont {Mrudul}}, \bibinfo {author} {\bibfnamefont {S.}~\bibnamefont {Mankovsky}},\ and\ \bibinfo {author} {\bibfnamefont {P.~M.}\ \bibnamefont {Oppeneer}},\ }\bibfield  {title} {\bibinfo {title} {Magnon orbital {{Nernst}} effect in altermagnets},\ }\href {https://doi.org/10.1038/s41535-026-00853-z} {\bibfield  {journal} {\bibinfo  {journal} {npj Quantum Mater.}\ }\textbf {\bibinfo {volume} {11}},\ \bibinfo {pages} {25} (\bibinfo {year} {2026})}\BibitemShut {NoStop}%
\bibitem [{\citenamefont {Huang}\ \emph {et~al.}(2026)\citenamefont {Huang}, \citenamefont {Wan},\ and\ \citenamefont {Tang}}]{HuangPrediction2026}%
  \BibitemOpen
  \bibfield  {author} {\bibinfo {author} {\bibfnamefont {L.}~\bibnamefont {Huang}}, \bibinfo {author} {\bibfnamefont {X.}~\bibnamefont {Wan}},\ and\ \bibinfo {author} {\bibfnamefont {F.}~\bibnamefont {Tang}},\ }\href@noop {} {\bibinfo {title} {Prediction of {{Magnetic Topological Materials Combining Spin}} and {{Magnetic Space Groups}}}},\ \bibinfo {howpublished} {arXiv:2601.05142} (\bibinfo {year} {2026})\BibitemShut {NoStop}%
\bibitem [{\citenamefont {Song}\ \emph {et~al.}(2026)\citenamefont {Song}, \citenamefont {Qi}, \citenamefont {Fang}, \citenamefont {Fang},\ and\ \citenamefont {Weng}}]{SongUnified2026}%
  \BibitemOpen
  \bibfield  {author} {\bibinfo {author} {\bibfnamefont {Z.}~\bibnamefont {Song}}, \bibinfo {author} {\bibfnamefont {Z.}~\bibnamefont {Qi}}, \bibinfo {author} {\bibfnamefont {C.}~\bibnamefont {Fang}}, \bibinfo {author} {\bibfnamefont {Z.}~\bibnamefont {Fang}},\ and\ \bibinfo {author} {\bibfnamefont {H.}~\bibnamefont {Weng}},\ }\bibfield  {title} {\bibinfo {title} {Unified {{Symmetry Classification}} of {{Magnetic Orders}} via {{Spin Space Groups}}: {{Prediction}} of {{Coplanar Even-Wave Phases}}},\ }\href {https://doi.org/10.1103/zy7s-j86r} {\bibfield  {journal} {\bibinfo  {journal} {Phys. Rev. X}\ }\textbf {\bibinfo {volume} {16}},\ \bibinfo {pages} {031038} (\bibinfo {year} {2026})}\BibitemShut {NoStop}%
\bibitem [{Mag()}]{Magnetic_Grp_Notes}%
  \BibitemOpen
  \href@noop {} {}\bibinfo {note} {Magnetic space groups are divided into four types. Type-I and type-II groups are ordinary space groups without and with time-reversal symmetry, respectively. Type-III and type-IV groups do not contain time reversal as an independent symmetry but include nontrivial combinations of time reversal with spatial transformations; the two types are distinguished by whether translations are combined with time reversal~\cite{bradley2010symmetry}.}\BibitemShut {Stop}%
\bibitem [{SSG()}]{SSGSupplement}%
  \BibitemOpen
  \href@noop {} {}\bibinfo {note} {See Supplemental Material for the spinor lifts, translation algebra, conjugation identities, removal of reciprocal-shift coboundaries, two-point-orbit construction, explicit platycosm realizations, and complete magnetic-group realizability table.}\BibitemShut {Stop}%
\bibitem [{\citenamefont {Zhang}\ \emph {et~al.}(2026)\citenamefont {Zhang}, \citenamefont {Wang},\ and\ \citenamefont {Zhao}}]{ZhangProjectiveRepresentation2026}%
  \BibitemOpen
  \bibfield  {author} {\bibinfo {author} {\bibfnamefont {Z.}~\bibnamefont {Zhang}}, \bibinfo {author} {\bibfnamefont {P.}~\bibnamefont {Wang}},\ and\ \bibinfo {author} {\bibfnamefont {Y.~X.}\ \bibnamefont {Zhao}},\ }\href {https://doi.org/10.48550/arXiv.2609.21289} {\bibinfo {title} {Projective representation theory of spin space groups}} (\bibinfo {year} {2026}),\ \Eprint {https://arxiv.org/abs/2609.21289} {arXiv:2609.21289 [cond-mat.mes-hall]} \BibitemShut {NoStop}%
\bibitem [{\citenamefont {Artin}(2011)}]{artin2011algebra}%
  \BibitemOpen
  \bibfield  {author} {\bibinfo {author} {\bibfnamefont {M.}~\bibnamefont {Artin}},\ }\href@noop {} {\emph {\bibinfo {title} {Algebra}}},\ \bibinfo {edition} {2nd}\ ed.\ (\bibinfo  {publisher} {Pearson Prentice Hall},\ \bibinfo {address} {Boston},\ \bibinfo {year} {2011})\BibitemShut {NoStop}%
\bibitem [{\citenamefont {Mackey}(1958)}]{MackeyUnitary1958}%
  \BibitemOpen
  \bibfield  {author} {\bibinfo {author} {\bibfnamefont {G.~W.}\ \bibnamefont {Mackey}},\ }\bibfield  {title} {\bibinfo {title} {Unitary representations of group extensions. {{I}}},\ }\href {https://doi.org/10.1007/BF02392428} {\bibfield  {journal} {\bibinfo  {journal} {Acta Math.}\ }\textbf {\bibinfo {volume} {99}},\ \bibinfo {pages} {265} (\bibinfo {year} {1958})}\BibitemShut {NoStop}%
\bibitem [{\citenamefont {Bradley}\ and\ \citenamefont {Cracknell}(2010)}]{bradley2010symmetry}%
  \BibitemOpen
  \bibfield  {author} {\bibinfo {author} {\bibfnamefont {C.~J.}\ \bibnamefont {Bradley}}\ and\ \bibinfo {author} {\bibfnamefont {A.~P.}\ \bibnamefont {Cracknell}},\ }\href@noop {} {\emph {\bibinfo {title} {The Mathematical Theory of Symmetry in Solids: Representation Theory for Point Groups and Space Groups}}},\ Oxford Classic Texts in the Physical Sciences\ (\bibinfo  {publisher} {OUP Oxford},\ \bibinfo {address} {Oxford},\ \bibinfo {year} {2010})\BibitemShut {NoStop}%
\bibitem [{\citenamefont {Bradley}\ and\ \citenamefont {Cracknell}(2009)}]{BradleyMathematical2009}%
  \BibitemOpen
  \bibfield  {author} {\bibinfo {author} {\bibfnamefont {C.}~\bibnamefont {Bradley}}\ and\ \bibinfo {author} {\bibfnamefont {A.}~\bibnamefont {Cracknell}},\ }\href@noop {} {\emph {\bibinfo {title} {The {{Mathematical Theory}} of {{Symmetry}} in {{Solids}}: {{Representation Theory}} for {{Point Groups}} and {{Space Groups}}}}}\ (\bibinfo  {publisher} {Oxford University Press},\ \bibinfo {address} {Oxford, New York},\ \bibinfo {year} {2009})\BibitemShut {NoStop}%
\bibitem [{\mbox{\emph{MG Computational Package}}()}]{MomentumSpaceSymmetryPackage}%
  \BibitemOpen
  \mbox{\emph{MG Computational Package}},\ \href@noop {} {}\bibinfo {howpublished} {\url{https://github.com/Peiyuan-Wang/SSGReciprocalDatabase/releases/tag/v0.12.1}}\BibitemShut {NoStop}%
\bibitem [{\mbox{\emph{MG Database}}()}]{MomentumSpaceSymmetryDatabase}%
  \BibitemOpen
  \mbox{\emph{MG Database}},\ \href@noop {} {}\bibinfo {howpublished} {\url{https://github.com/Peiyuan-Wang/SSGReciprocalOfflineDatabase/releases/tag/v0.1.5}}\BibitemShut {NoStop}%
\end{thebibliography}%

\clearpage
\onecolumngrid
\begin{center}
\textbf{End Matter}
\end{center}
\twocolumngrid
\appendix
\section{Appendix: Computational package and database}
\label{app:software}

\subsection{Determining the MG of an SSG}
For each SSG in Ref.~\cite{XiaoSpin2024}, the calculation is practiced through the following four steps.
\begin{enumerate}
\item Determine the subgroup $T_0$ or $Z_{T_0}$ of translations, and therefore the BZ as well as the reciprocal lattice $L_F$.
\item Derive the $P$-action on the BZ using
Eqs.~\eqref{eq:M-Action} and \eqref{eq:Central-Result}.
\item Determine the cohomology class of $\bm{\kappa}$ using the Smith-normal-form method (see the SM~\cite{SSGSupplement}), i.e., test whether a change of momentum origin can remove all fractional reciprocal translations. 
\item Identify the MG from the $P$-action on the BZ and the unitarity grading $s_p$.
\end{enumerate}
MG names and numbers follow the Belov--Neronova--Smirnova (BNS) convention~\cite{BradleyMathematical2009}. The catalog covers 67,475 SSG labels in
Ref.~\cite{XiaoSpin2024}; families with continuous spin parameters are
represented by selected parameter choices.

\subsection{Mathematica program package}
The \texttt{SSGReciprocalDatabase} Mathematica package can be downloaded
from Ref.~\cite{MomentumSpaceSymmetryPackage}.
Download the \texttt{.paclet}
file, then install and load it in Mathematica by
\begingroup\small
\begin{verbatim}
PacletInstall[
 "SSGReciprocalDatabase-0.12.1.paclet"]
Needs["SSGReciprocalDatabase`"]
\end{verbatim}
\endgroup
Replace the filename by the path to the downloaded file.
Enter an SSG label in Ref.~\cite{XiaoSpin2024} as a string, for example \texttt{"N6.9.19"}.
The main functions include the followings.
\begin{itemize}
\item \texttt{getSSGZL0[ssg]} gives the subgroup $T_0$ or $Z_{T_0}$ of translations that
defines the BZ as well as the reciprocal lattice $L_F$.
\item \texttt{showSSG\allowbreak Reciprocal\allowbreak GenTab[ssg]} displays the real-space
generators of $P$ and its action on the BZ, by providing the fractional reciprocal
translations $\bm\kappa_p$, the unitarity grading $s_p$,
and the grading $\eta_p$ when
a common spin axis exists.  Use \texttt{getSSG\allowbreak Reciprocal\allowbreak GenTab[ssg]} for
the same information in a form suitable for further Mathematica calculations.
\item \texttt{getSSG\allowbreak Nonsymmorphic\allowbreak Info[ssg]} reports
whether the MG is an MNG.
\item \texttt{getSSG\allowbreak Reciprocal\allowbreak Magnetic\allowbreak SpaceGroup[ssg]} gives the MG number and type I--IV.
\texttt{showSSG\allowbreak Reciprocal\allowbreak MagneticGroup[ssg]} displays its generators
alongside the standard MG operations.  The corresponding ordinary group,
obtained by forgetting antiunitary labels, is given by
\texttt{getSSG\allowbreak Reciprocal\allowbreak Family\allowbreak SpaceGroup[ssg]}.
\item \texttt{getSSGLabelsForBNS[bns]} runs the search in reverse: its
input is a quoted MG number such as \texttt{"7.24"}, and its output is
the complete list of SSGs in the catalog that realize the MG.
\end{itemize}
\texttt{getSSG\allowbreak ReciprocalData[ssg]} provides the
complete momentum-space symmetry data for each SSG, \texttt{SSGGraded\allowbreak Reciprocal\allowbreak GroupElements[ssg]} lists the
full momentum-symmetry operations, and
\texttt{SSGO3\allowbreak Representation\allowbreak Matrices[ssg]} gives the spin-space matrices.
\texttt{SSGRecompute\allowbreak ReciprocalData[ssg]} permits a selected representation
or parameter choice to be calculated separately.  



\subsection{Offline database}
The offline database is available for macOS (Apple Silicon) and Windows (64-bit) from
Ref.~\cite{MomentumSpaceSymmetryDatabase}.
For macOS, download the database file and viewer, unzip the viewer, and
place the database beside the application.  For Windows, extract the
archive and open \texttt{SSGReciprocalDatabase.exe}.

In \texttt{SSG lookup}, enter an SSG label in Ref.~\cite{XiaoSpin2024},
such as \texttt{N6.9.19}, to view its magnetic-order type, $T_0$ or $Z_{T_0}$, and associated MG. The generator panels display
the $O(3)$ spin matrices and the $P$-action on the BZ, including the fractional reciprocal translations $\bm\kappa_p$ and the
unitarity grading $s_p$. The group panel compares the calculated
operations with the standard MG setting.

The reverse lookup accepts an MG number, such as \texttt{7.24}, and
lists the SSGs that realize it, together with their magnetic-order types and corresponding magnetic space groups.

\end{document}